Technische Universität München

Master Thesis

# Decentralized Synchronization
# for Wireless Sensor Networks

*Author:*
Hauke HOLTKAMP

*Supervisor:*
Dr. Robert VILZMANN

arXiv version

March 31, 2008

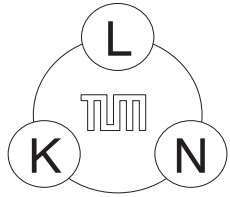

**Technische Universität München**
**Lehrstuhl für Kommunikationsnetze**
Prof. Dr.-Ing. Jörg Eberspächer

# Master Thesis

## Decentralized Synchronization
## for Wireless Sensor Networks

Hauke Andreas Holtkamp

# Abstract

**Decentralized Synchronization for Wireless Sensor Networks**

Due to their heavy restrictions on the hardware side, Wireless Sensor Networks (WSN) require specially adapted synchronization protocols to maximize measurement precision and minimize computation efforts and energy costs. A promising approach is given by the "Firefly Protocol". Inspired by the behavior of fireflies it is intrinsically robust, specific to the wireless broadcast nature of WSNs and promises high precision. So far only theoretically evaluated, this thesis implements the "Firefly Protocol" on a system of MICAz Berkeley motes using TinyOS 2.x. In order to implement the theoretical framework on actual hardware, several adaptations were made to compensate hardware delays. Although Berkeley motes have the advantage of being readily available and highly flexible, they bear many delay sources which have to be addressed. In small networks, the protocol was found to deliver precisions up to three microseconds over one hop.

# Contents

# Chapter 1

# Introduction

**Characterization of Wireless Sensor Networks**

Diminishing prices in electronics and increasing overall computing capacity enable the creation of small and cheap sensors, which allow to monitor the environment in detail. For example, even the simple and well established application of a fire detection system in a large residential building would be greatly reduced in price and complexity if it were possible to use ad hoc network nodes with built-in smoke sensors instead of wired detectors. Being able to measure the "real world" efficiently creates new options for future computer applications in the form of **Wireless Sensor Networks** (WSN). WSNs consist of low-cost, low-power sensor nodes that are small in size and create a multihop network through radio communication over short distances. Such networks possess numerous differences as compared to currently available wireless systems. Since wireless sensor nodes run on batteries, energy efficiency is the prime target to maximize life time, as batteries' sizes and capacities will not change dramatically over the coming years (1). Compared to current systems like mobile phones or wireless LAN, sensor lifetimes are aimed significantly higher in the range of several years. To enable small and energy efficient network nodes the system has to be ad hoc in nature and communicate over multiple hops instead of a base station cell setup (like today's mobile phone networks). To save energy, nodes should only actively communicate with the network when they have data to transmit or are queried.

Due to pricing, there are severe limits on the available hardware in a sensor node. Currently available sensor nodes have the computation and storage capacities of personal computers from the 1980s (e.g. an 8 MHz processor). The clock (time) precision is limited by the internal quartz quality, which is a determining factor in synchronization. In addition, depending on the type of application the sensor nodes may become so cheap, that failure rates go up and, combined with lack of battery power, frequent node losses can be expected and have to be accounted. It is very attractive to imagine WSNs with hundreds or even thousands of nodes, all gathering data over a large area. This opens challenges regarding scalability and high node densities and precludes manual configuration of individual nodes (2). A major difference to all existing computer networks (e.g. the internet) is that WSNs gather sensing data. Instead of high bandwidth or computation power, WSN system designs aim at taking timed measurements. This type of network is called *data centric* (3) (as compared to *network centric*) because the focus is on gathering data with specific properties such as, for example, location, time or type.

## Applications

Over the past years the first WSN systems have been deployed. After introducing some particularly interesting examples, an outlook into possible extended scenarios is given below

**Ocean Water Monitoring** In the ARGO project[1] free-drifting sensor units are used to observe temperature and salinity in the upper level of earth's oceans for climate analysis. Once dropped from ships these nodes cycle through different ocean depths. Whenever they are at surface, the nodes report the measurement data via satellite. As of February 2008, over 3000 nodes were actively deployed, each costing more than US$ 10,000.

**Parking Space Localization** Many of today's parking garages have wired car sensor at each parking space to accurately determine the number of free spaces and direct incoming cars to their destination quickly. Project Networked Parking Spaces (4) uses WSNs which form a static multihop network to achieve the same goal. Even cars may be equipped with sensor nodes to query the system for free spots.

**Acoustic Localization** For a posteriori clues, law enforcement (5) uses multihop WSNs to localize a sound source, like a sniper. For example, sound sensors may be deployed around a podium at a president's speech. By comparing the time of arrival of a gunshot sound at each sensor node, a sniper can be localized with a precision of about one meter.

Outlook - Sensor networks can be envisioned in almost any area of life, be it species monitoring, environmental monitoring, agriculture, production and deliver, disaster relief, building and automation, traffic and infrastructure, home and office, healthcare or military and law enforcement (3). An entire hospital may equip its patients with sensors to observe their vital signs without inhibiting their mobility. Also, the military is specifically interested in WSNs for battlefield monitoring in order to quickly establish a well supervised perimeter by dropping sensors from an airplane and to decrease the number of personnel in the danger zone.

---

[1]ARGO - Global Ocean Sensor Network. www.argo.ucsd.edu

# Chapter 2

# Synchronization in Wireless Sensor Networks

## 2.1 Synchronization Aspects

This thesis implements a new type of synchronization protocol. In order to evaluate a synchronization protocol, it is important to understand the critical factors of timing.

In WSNs, energy efficiency is the prime design goal. Sleep modes (where almost all parts of a node are switched off except for the internal clock or a low power listening component) allow to reduce power consumption to less than 0.1 percent of the power during transmission[1]. Therefore, it is of major importance to synchronize sleep modes to perform concurrent measurements and to be able to communicate over multiple hops. Take the scenario of a node in a wireless network, which wakes up to transmit data, but finds that all other nodes are still in sleep mode. The node would then have to remain powered up waiting for other nodes to awake, thus wasting precious energy. Low power sensing is an option to wake a sleeping node. However, precise and reliable synchronization is by far the superior option to choose, since it can be used for additional tasks like precise measurements.

Unlike centralized systems, where there is no time ambiguity and a clear ordering of events, distributed systems (like a sensor network) have no global clock or common memory. Each internal clock has its own notion of time which may easily drift seconds per day, accumulating significant errors over time (6).

Two types of synchronization exist, *(global) time synchronization* and *slot synchronization*(7; 8): In a globally time synchronized WSN, all members are aware of the current time on a common timescale. This allows to timestamp measurement data at the time of its taking. Like "This reading was taken on June 1st, 2008 at 12:10 pm and 235 ms". Alternatively, some synchronization systems allow to recursively conclude at what time data was taken at some later point in time. Here, the node does not know the date or time, but it can estimated backwards at the time of data collection by the collector.

---

[1]MICAz Datasheet:
http://www.xbow.com/Products/Product_pdf_files/Wireless_pdf/MICAz_Datasheet.pdf

Slot synchronization, on the other hand, ensures that all nodes have a common perception of time frames (i.e. slots). The borders of these slots are matched exactly, such that they can be subdivided or used as a time unit. For example, a node may be instructed to receive or take a measurement at the beginning of every slot. If nothing is perceived, they go back to sleep. These slots can have arbitrary length from milliseconds to hours or even days.

Often, the limiting factor on synchronization (as on almost any engineered device) is the production cost. Many applications only become feasible, if their cost is below a certain threshold (e.g. (1) proposes US$ 1 for a standard sensor node). If money was not an issue, all network nodes would be equipped with a costly GPS receiver. However, only in few systems (e.g. Ocean Water Monitoring), GPS is feasible due to the fact that other cost factors outweigh the GPS receiver price[2].

In high volume low-price sensor nodes this is not an option. Here, the internal quartz has to be used which introduces drifts. Temperature, supply voltage, crystal impurity, pressure, etc. all have an influence on this drift. To make sure that the clocks in the network do not drift too far apart, they need to be matched regularly. This matching is the synchronization process. At some point, every synchronization system designer has to make a tradeoff between the maximum reachable precision and the communication overhead (see Figure 2.1). More frequent alignments allow to better compensate the clock drift, but require extra message exchange which blocks the channel and uses precious energy. If synchronization is only performed in distant intervals, however, the individual clocks will inevitably drift and reduce reliability and precision.

In the following section, some popular synchronization protocols are introduced which attempt to fulfill these requirements.

## 2.2  Literature Survey

This section introduces the most prominent synchronization protocols for wireless sensor networks in chronological order. They are not yet aimed at specific applications but try to maximize the synchronization precision for ad hoc multihop networks. First, four well established protocols are introduced in detail. Second, three protocols are described, which make a novel approach to reach synchrony in a network.

---

[2]In addition to the price, GPS is not feasible for indoor applications due to the line of sight condition to the satellites.

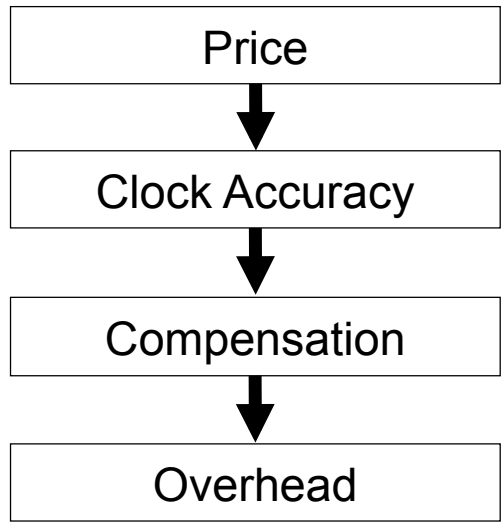

Figure 2.1: The production price determines the possible inherent clock precision. A less precise clock requires more frequent compensation. Compensation creates communication overhead which is costly in regard to bandwidth and energy.

### 2.2.1 Prominent Synchronization Protocols

Wireline synchronization protocols like the Network Time Protocol (NTP) (9; 10) are not applicable to WSNs due to their large memory requirements and comparably low precision (10 - 100ms). Therefore, in recent years several protocols have been proposed to create synchrony between self-organizing network nodes, each of them tackling specific problems arising in ad hoc networks, reaching precisions of up to 2.24 $\mu$s per hop (11; 12). However, high precision comes with a trade-off in complexity and scalability. Whilst designed for and tested in numbers of up to 60 nodes (12), it is yet open how these approaches will perform in numbers of several hundreds or thousands. Most of these established protocols have been around for several years and have been tested and described in detail. These are the most prominent ones[3]:

#### Synchronization by Römer

Kay Römer's approach from 200 (17) tried to address different challenges than many of the later to come synchronization protocols. Not aiming at a high precision, Römer developed a technique to compare temporal relationships and realtime issues in sparse ad hoc networks. The focus was for nodes to be able to communicate their measurements with timestamps *a posteriori*, even if at the time of the real world occurence the nodes were not in communication proximity, for example in an inherently mobile sensor environment or a drive-by data collection setup. The protocol is not intended to synchronize all nodes of the network to a common time. Instead, it only adjusts the timestamp of a message locally as it is passed to its destination. This is achieved through a round trip delay measurement on each hop. The achieved maximum precision of 3 ms (over 5 hops) is only sufficient for few applications. Nevertheless, Römer's work was one of the early practical approaches and was interpreted and improved by many synchronization protocols to follow.

---

[3]Several additional synchronization protocols have been suggested (13; 14; 15; 16) which explore different aspects of synchronization in wireless networks, but are not directly connected to this work.

**Reference Broadcast Sychronization (RBS)**

In 2002, Elson and Estrin (11) proposed the widely regarded Reference Broadcast Synchronization (RBS) which was tailored for sensor network requirements and adressed system specific delays in order to maximize precision, reaching 3 $\mu$s. The two major innovations were the use of reference messages and the elimination of the non-deterministic hardware delays *Send Time* (propagation of a message through the OSI layers) and *Access Time* (time to wait until the channel is free). In RBS, any node can send a reference beacon signal. All adjacent nodes who pick up this beacon timestamp it and then compare their findings with each other in order to adjust their internal clocks. Over time, all nodes will have referenced and compared their findings and the network will synchronize. The main strength of RBS is its applicability to commodity hardware and existing software in sensor networks.

**Timing-sync Protocol for Sensor Networks (TPSN)**

The Timing-sync Protocol for Sensor Networks (TPSN) (18), 2003, was a direct reply to RBS. Ganeriwal et al. specifically chose a strongly hierarchical approach, combined with pair-wise synchronization, while using some of RBS' suggestions like reduction of the critical hardware delays. At first setup, a tree structure is established amongst equal nodes which from then on is the basis for top-down pair-wise synchronization. During each synchronization hop the round trip time is measured. The constant updates from the surrounding nodes are used for correction of the internal clock. While TPSN argues to be *theoretically* superior to RBS by a factor of 2 in terms of accuracy, the provided laboratory test results are in the range of 17 $\mu$s which is below the 3 $\mu$s accuracy of RBS. The carefully established tree structure setup has not been tested towards robustness or mobility and appears to be a rather complex approach to synchronization. The authors did not go into detail how the protocol could support a sensor sleep mode without breaking the tree structure. It is therefore to be seen how well TPSN synchronization will converge, perform and deliver in real world applications. However, it is noteworthy that TPSN was simulated in an environment of 300 nodes. This is the largest test provided in a WSN synchronization publication.

**Flooding Time Synchronization Protocol (FTSP)**

Maroti et al. (12) proposed the Flooding Time Synchronization Protocol (FTSP) in 2004, utilizing previous progress from RBS and TSPN. FTSP combines a dynamic flat node hierarchy with medium access control (MAC) layer timestamping and clock correction. First, the node with the lowest ID is elected as master. The master node regularly floods the network with timing information. Eliminating most possible hardware delays the nodes communicate single timestamps (no round trip information) through the network. The internal clocks are continually updated and corrected via linear regression. Remarkably, the authors ran a very large live experiment on 60 network nodes, testing on malicious devices, mobility and root election. Here, the average clock offset was 2.24 $\mu s$ (maximum 8, 64 $\mu s$), capping RBS. Also, this is the only approach for which the convergence time[4] (15 min equalling 30 synchronization rounds) is listed which can be a determining criterion for the choice of a sychronization protocol.

## 2.2.2   Novel Synchronization Protocols

Whilst the previously described protocols use well-known procedures from computer networking, all of the following go beyond the more traditional attempts in synchronization and try to exploit the cooperative nature of wireless networks in more detail. Here, the advantage is that theoretically the synchronization precision does not degrade over the number of hops, it is reached through a consensus amongst nodes. As can be seen in a closer study, it is difficult to compare these suggestions with the above, since they are strongly theoretical and (in their original descriptions) require different hardware than standard motes (apart from Tyrrell et al. (19)).

This novel suggestion comes from biology and is described in detail in Chapter 3.1. Inspired by the phenomenon of thousands of fireflies gathering on trees in Malaysia and pulsing in synchrony, the approaches assume a very high number and density of nodes. These protocols have only been simulated and not yet been technically implemented.

**Scalable Synchronization Protocol**

Inspired by the challenge of scalability and nature's example for a solution in the form of Malaysian Fireflies, Scaglione et al. (20) propose a protocol for slot synchronization with minimal message overhead in 2005. Employing the PCO model (21) all nodes slowly adjust to a pulse. The authors prove in theory that the nodes will quickly yield a common time scale. Although making some assumptions regarding actual hardware and delays, this model is still very theoretical and makes no specification regarding precision. Simulations suggest that the convergence time is optimally short. Scaglione et al. suggest to apply this system for a waveform "reach-back", where data can be transmitted over a longer distance by the sensor group than each individual would have accomplished, for example through pulse-position modulation. A strong practical drawback stems from the fact that this approach requires specific hardware that emits and senses the synchronization pulse solely on the physical layer (i.e. no software interpretation).

---

[4]Convergence Time: Time to reach synchrony after startup

**Algorithmically Optimal Time Synchronization**

Servetto et al. (22) propose a protocol that is very closely related to Scaglione et al. (20) and shares most of the ideas and conclusions. The two authors do not refer to each other and have published in close succession. Picking up the problem of scalability in sensor networks, Servetto et al. suggest an approach where an elected and centrally located node initiates regular pulses and data packets that propagate through the network. The surrounding nodes listen to the pulses and messages and "tune in". In this way, a synchonization waveform is established throughout the network. The election procedure for the master node is not provided. Like Scaglione et al., Servetto et al. suggest to apply this method to reach back to a distant receiver which can either be synchronized or obtain sensor data. Although the protocol proposes some innovations like node density dependent power scaling, it is nearly identical to the Scalable Synchronization Protocol.

**Firefly Role Model Synchronization**

The most recent and most practical approach of the "Firefly Models" is by Tyrrell et al. (19; 23). Based on Scaglione et al.'s findings, it transforms the model to transmission messages of non-zero length (i.e. no pulses) and works with some assumptions on delays that can occur in hardware. Also, node deafness during transmission is considered which poses a major difference between off-the-shelf hardware and the theoretical model. This approach is the basis for this work and is described in detail in Chapter 3.

## 2.3 Summary

Table 2.1 summarizes all of the described protocols regarding key aspects.

**Master/Slave or P2P**  Hierarchy is an important factor for the applicability of a protocol. Do all nodes behave the same or does a master node exist? Whereas TPSN creates a strict tree hierarchy which make is sensible to topology changes, protocols like by Scaglione et al. or Tyrrell et al. assume all network nodes to be equal members of a peer-to-peer system.

**Clock Correction**  For maximum accuracy it is indispensable to compensate for the inherent clock drift by internal clock correction in each network node.  Although some protocols do comment on this point, this survey states which ones have actively used clock correction to achieve higher precisions.  Clock correction allows to space communication intervals further apart, as each node "learns" about its long term drifts.

**Internal/External Synchronization**  Protocols strongly differ regarding their type of time reference.  Whereas RBS, TPSN and FTSP explicitly aim at distributing a (global) external reference time into the network, the other approaches only ensure that the clocks inside the network are synchronized, disregarding the external time frame.

**MAC Layer Control Necessity**  For a wider usability on standard hardware, it is desirable for a protocol to have little to no requirements regarding the medium access control or physical layer.  For timestamping it is necessary to have limited access to the packet queue in the radio chip, which several protocols require.  The novel protocols, however, assume that extra radio hardware is available, aside from the standard radio components.

**Maximum Clock Offset**  This column reflects the maximum precision that each protocol claims. Since some have only been simulated so far, no values are provided.

**Tested Network Size**  Since scalability is a highly anticipated criterion in WSNs, it is displayed here, which network sizes (i.e. number of nodes) the authors have tested their system on, if available.

**Robustness**  Since node failures and topology changes can be expected in most scenarios, this column is an estimate of the protocol robustness by the author of this work.  TPSN receives a low rating due to the strict hierarchy of the approach. All P2P protocols can be expected to possess high robustness due to their inherent self organization.

**Convergence Time**  Although this is a very interesting parameter, it is only provided by two protocols.  It describes the time a network requires to reach synchrony (starting from a random setting). Since comparison intervals differ, the duration is given in the unit of "synchronization periods $T$". For example, in FTSP $T = 30\ s$.

**Synchronization Type (Slot/Global)**  The protocols can be split into two groups (plus one special case) regarding the type of time they provide to the network. Either they provide a global reference (e.g. Universal Time Coordinated (UTC)) or create time slots of predefined lengths. Römer's suggestions focuses on the causal ordering of events more than the exact time and is marked specifically.

**Intended Node Density** Synchronization protocols are specifically aimed at certain densities. Sparse networks require a protocol to be robust and be able to reliably distribute time over a single and unreliable link. High density network protocols use the multiple existing paths and broadcast nature to their advantage. When a protocol for sparse networks is deployed in a high density environment, contention may occur due to too many messages occupying the channel. A high density protocol in a sparse network may not find the minimum number of nodes in range necessary to perform a synchronization step. Römer, for example, aims at nodes which may be out of reach for a long time. In contrast, Servetto et al. assumes that all nodes are in close proximity, if not even in single hop range.

**Targeted/Broadcast Transmission** It is interesting for the classification of a protocol whether it uses the broadcast property of the wireless channel or whether it behaves like in a wired system. For example, TPSN specifically addresses each node and only performs round trip measurements. Other protocols strongly base on the fact that a message will always be heard by multiple receivers.

**Message Overhead** For energy efficiency, it is of prime importance how much radio communication is necessary to keep the network synchronized. This column is an estimate of the author, interpreting protocol descriptions. RBS receives a high overhead rating, because nodes always have to compare their perceived reference with all other nodes in range, which may become a large number. Scaglione et al., in contrast, employs special pulsing hardware which does not create message overhead.

| Protocol / Feature | Master/Slave or P2P | Clock Correction | Internal/External Synchronization | MAC Layer Control Necessary | Maximum Clock Offset | Tested on Network Size of $N$ Nodes | Robustness | Convergence Time | Synchronization Type (Slot/Global) | Intended Node Density | Targeted/Broadcast Transmission | Message Overhead |
|---|---|---|---|---|---|---|---|---|---|---|---|---|
| Römer | P2P | No | Internal | No | 3ms | - | High | Unkn. | Event[a] | Low | Target | Low |
| RBS | P2P | No | Both | Yes | $3\mu s$ | 2-20 | Med | Unkn. | Global | Med | Broadc | High |
| TPSN | M/S | No | Both | Yes | $17\mu s$ | 150-300[b] | Low | Unkn. | Global | Any | Target | Low |
| FTSP | P2P | Yes | Both | Yes | $2.24\mu s$ | 60 | High | 30 T | Global | Med. | Broadc | Med |
| Scaglione | P2P | No | Internal | PHY[c] | Unkn. | - | High | Unkn. | Slot | Any | Broadc | None |
| Servetto | M/S | No | Internal | PHY | Unkn. | - | High | Unkn. | Slot | High | Broadc | Low |
| Tyrrell "Firefly" | P2P | No | Internal | Yes | Unkn. | - | High | 32 T | Slot | Low | Broadc | Med |

Table 2.1: Comparison of Synchronization Protocols

[a] Focuses on the order of events more than the timestamps
[b] Only simulated
[c] Requires specific hardware that operates solely on the physical layer, e.g. a pulse generator

# Chapter 3

# Firefly Synchronization

The present chapter introduces the biological phenomenon which inspired researchers to scrutinize a novel approach in synchronization. After following the research that has led to the Firefly Protocol, the necessary adaptations are discussed which were made in this thesis.

## 3.1 Biological Inspiration

Several species of Malaysian fireflies (e.g. Pteroptyx Malaccae in Figure 3.1) exhibit the stunning phenomenon of reaching synchrony while flashing. Male fireflies gather on trees, randomly flashing once per second. Over time, synchrony emerges without relying on a central entity. To understand this process, a set of experiments was conducted by Buck et al. (24) in 1981. By external excitation, Buck was able to influence the natural flashing period of 965 ms ± 90 ms. He found out, that the firefly would not react to signals that occur right after the insects natural flash. However, excitations that occurred after this *refractory period* would lead the firefly to adjust its natural period to an earlier time, thus trying to reach synchrony. This refractory period is the key element that allows oscillators to reach synchrony in an apparently chaotic environment (see Figure 3.2).

Regarding the purpose of firefly synchronization, biologists still disagree, especially because the majority of firefly species does not exhibit this behavior. Theories include that synchrony serves as a noise-reduction mechanism or as a cooperative attraction on female fireflies (25).

The pulse-coupled synchronization phenomenon is not exclusive to fireflies and occurs in numerous other populations of oscillators. Examples include pacemaker cells in the heart (26), crickets chirping in unison (27) and women whose menstrual periods become mutually synchronized (28).

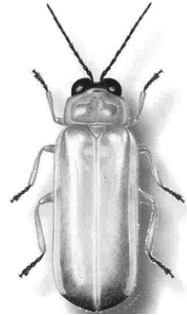

Figure 3.1: Pteroptyx Mallaccae (Source: www.rspg.org)

In 1990, Strogatz et al. (21) derived and proved a mathematical model for this phenomenon, called the *Pulse-Coupled Oscillator (PCO)*, which is described in detail in the following section.

## 3.2 Pulse Coupled Oscillators (PCOs)

Inspired by a model for two interacting oscillators in cardiac pacemakers (29), Mirollo and Strogatz (21) extended the model to entire populations of oscillators (like fireflies), by simplifying the dynamical phase function which describes the firing period. The internal clock of a firefly which determines the firing instant is modeled as an oscillator which interacts with other oscillators through discrete events (i.e. pulses).

Peskin describes the dynamics as

$$\frac{dx_i}{dt} = S_0 - \gamma x_i, \quad 0 \le x_i \le 1, \quad i = 1, ..., N.$$

When $x_i = 1$, the $i$th oscillator "fires" and $x_i$ jumps back to zero, for an initial condition $S_0$, dissipation $\gamma$ and $N$ oscillators.

Strogatz et al. lift the differential equation condition by assuming that $x$ will increase monotonically and smoothly from 0 to 1, described by a phase function $\phi_i(t)$. When $\phi_i(t)$ reaches 1, the oscillator "fires" and $\phi_i(t)$ is reset to zero. If not coupled to other oscillators, it will naturally fire with period $T$.

When coupling occurs (i.e. two oscillators fire in each others' range), the phase function will be adjusted as follows:

$$\begin{cases} \phi_j(\tau_j) = 0 \\ \phi_j(\tau_j) = \phi_i(\tau_j) + \Delta\phi(\phi_i(\tau_j)) \text{ for } i \ne j \end{cases}$$

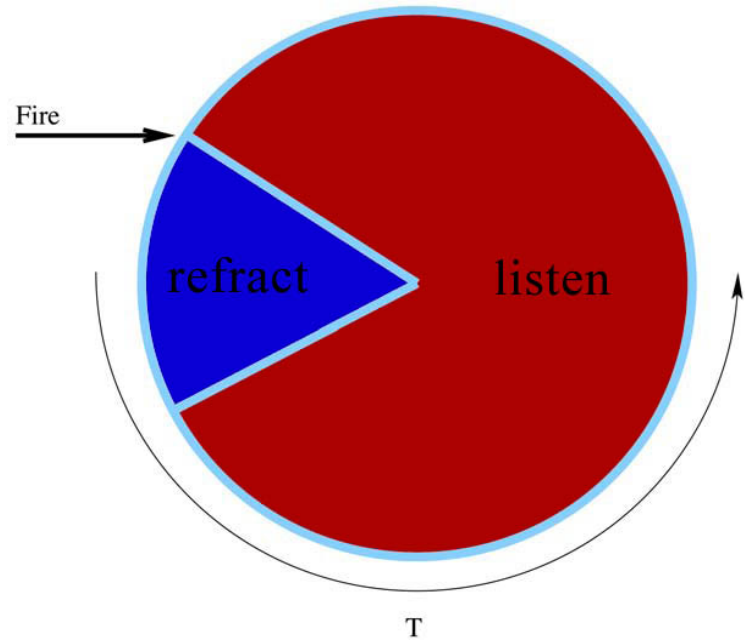

Figure 3.2: A simple schematic of the firefly's flashing period. After firing a flash of light, the insect is blind to incoming flashes for a short duration (refract phase). While in the listen phase, the firefly will try to adjust its firing points to reach synchrony.

Figure 3.3 plots the time evolution of the phase when receiving a pulse. The received pulse causes the oscillator to fire early. Using considerations of monotonicity and concavity, Strogatz et al. were able to show that by appropriate selection of $\Delta\phi$ a fully meshed network is able to synchronize within a few periods. The proof itself is out of the scope of this work and can be found in (21). Note that this model only constitutes the basis for Firefly Synchronization in networks nodes. Some assumptions like the communication through pulses do not hold and some aspects have been extended by Tyrrell et al. (19) and the author of this thesis.

The treatment of network nodes as PCOs has several advantages over a traditional point-to-point network perspective:

- It makes use of the inherent broadcast nature of wireless networks.

- It has been shown to have an asymptotically good settling time (20).

- It is intrinsically robust, as all nodes are truly equal and no hierarchy exists.

- It theoretically scales extensively and improves with increased node number and density (30).

Some drawbacks exist:

- Fireflies communicate through light which does not require a medium access scheme.

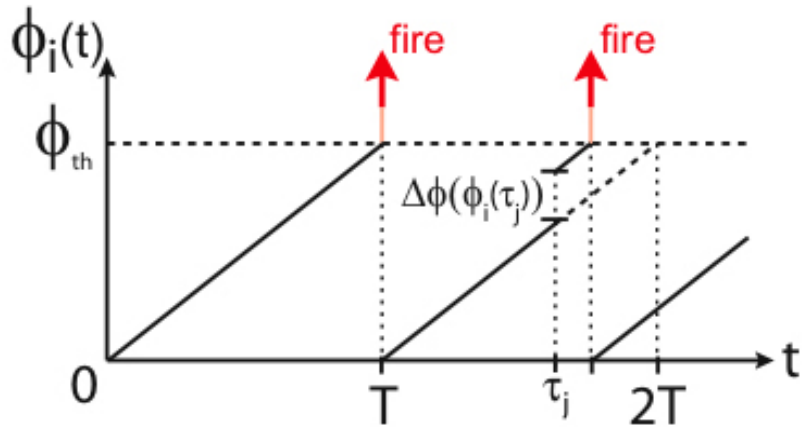

Figure 3.3: Time evolution of the phase function in the Firefly Protocol (23)

Wireless networks, on the other hand, rely on carrier sense multiple access (CSMA). This imposes a strong limitation on precision and scalability.

- Compared to most other sensor synchronization protocols this approach synchronizes slots instead of a universal time reference. Some sensor applications may require a universal time frame.

## 3.3   Application to Wireless Networks - The Firefly Protocol

The PCO model (see Chapter 3.2) has inspired numerous research groups (19; 22; 31; 20) to look into possible applications in communication networks. The most interesting being by Tyrrell et al. (23), because it tackles specific challenges common to wireless networks and is closest to an implementation on commodity hardware like MICAz.

Three issues are specifically addressed that allow utilization of the PCO model in network nodes:

- Node deafness during transmission

- Non-zero message length

- Processing delay

**Node Deafness**   Network nodes are usually equipped with a single radio chipset that can either send or receive with a tuning interval in between the two states. Therefore, a node is "deaf" while it transmits a message. If the PCO model were directly applied to network nodes the attainable precision would be lower bound by the message transmission time $T_{TX}$. While a node is transmitting, it is unable to recognize or interpret another node's synchronization message. For example, at 250 kbits/sec an 18 byte message would take 576 $\mu$s to transmit. This limit on precision is clearly unacceptable.

To overcome this limitation, Tyrrell et al. suggest to (randomly) create two groups in the network, each transmitting and listening in turn. Therefore, group A would be transmitting while group B listens and group A can then adjust its timing when B transmits.

**Non-Zero Message Length** Pulses (as suggested by Scaglione et al. (32)) are not a good option in radio communication systems since they require special hardware and are virtually impossible to detect. Rather, a synchronization word is used, which can be recognized by all receivers. Since a synchronization message has non-zero length, the PCO model has to be adapted. Tyrrell et al. suggest to reserve a fixed amount of time for the transmission of radio signals.

**Processing Delay** Similar to the finite message length problem, the PCO model has to be modified since digital receivers require significant encoding and decoding time. Maroti et al. (12) set these delays at around $200\mu$s. To allow the system to still synchronize, these delays have to be accounted for.

## 3.4 Phase Diagram

Combining these three challenges results in the Firefly phase diagram (Figure 3.4), which creates two groups of transmitters and reserves a transmission window $T_{TX}$.

**Single Node** When a node with Firefly Synchronization is switched on it will oscillate with period 2 T, starting at the point "Fire 1". It will run through the *refract* state, where the node ignores all synchronization influences. During the succeeding *listen* state, the node is susceptible to synchronization signals. If no messages are received, it will move into the *wait* state, which is similar to *refract*. At the end of the *wait* state, the node will internally issue the firing command. Because of the processing delay, this has to occur exactly on time such that the message will leave the transmitter at point "Fire 1". Therefore, according to Tyrrell et al., the *transmit* state is as long as the internal processing time of the node. After the message has left the transmitter at point "Fire 1", the cycle will reiterate.

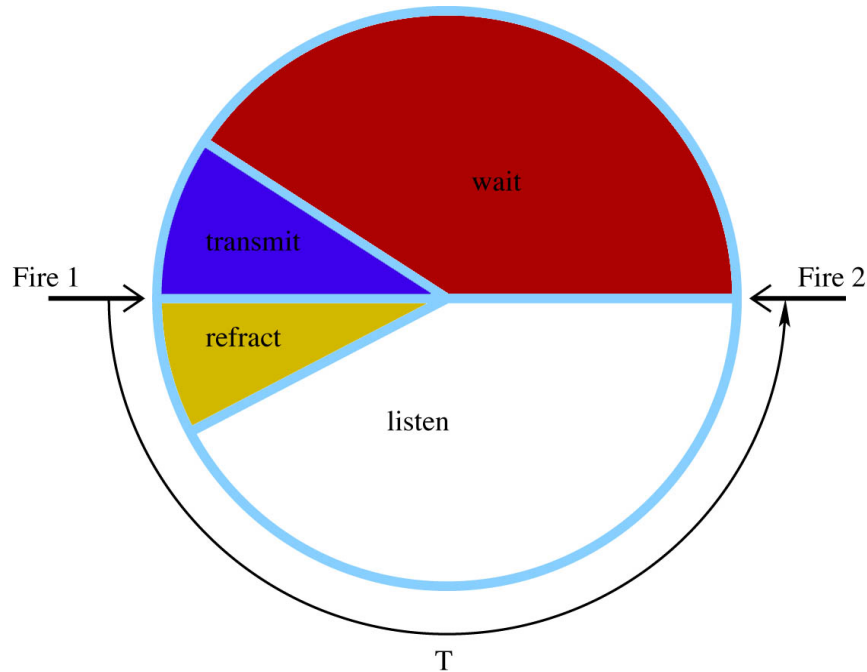

Figure 3.4: Phase diagram for an oscillator by Tyrrell et al. Two groups of oscillators form, spaced T apart.

**Multi-Node** Multiple nodes will influence each other. During the *refract* state, all received synchronization messages will be ignored by the node. However, during the *listen* state, upon receiving a "flash" or a synchronization message, respectively, it will adjust its phase towards point "Fire 2" in the form of an immediate phase jump. The phase adjustment brings the firing points of all nodes closer together. The *wait*, *transmit* and *refract* states remain unmodified. Since nodes will only react during the *listen* state, two groups form, spaced exactly T apart. When group A fires, group B is listening and vice versa. The membership of the groups is entirely random and only the depends on the time of entry into the network. In the worst case, all nodes will (from the start) be in one group. Then, they will not receive any synchronization messages during the *listen* state and the precision becomes capped by $T_{refract}$. A solution to this problem is suggested in this work.

Tyrrell et al. have simulated this setup to determine the optimal duration for each state. The best synchrony rate is achieved for $T_{refract} = 0.4 * T$. $T_{transmit}$ is given by the hardware configuration and estimated at $T_{transmit} = 0.1 * T$. Accordingly, $T_{wait} = 0.9 * T$ and $T_{listen} = 0.6 * T$. In this thesis, with a base period $T = 1$ s the duration of packet $T_p = 576$ $\mu$s, $T_{transmit} = 0.1 * T = 100$ ms is chosen. This is sufficient to fit 173 synchronization messages back-to-back which is safe enough for all considerations in this work (Table 3.1).

| State | wait | transmit/TX | refract | listen |
|---|---|---|---|---|
| Duration [ms] | 900 | 100 | 400 | 600 |

Table 3.1: State durations used in this work

## 3.5   Modifications of the Firefly Model in this Work

Although the Firefly Protocol introduces some powerful ideas that bring the PCO model closer to actual hardware, there are several significant weaknesses which have to be countered, before an implementation on network nodes is possible.

In order to implement Firefly Synchronization in a functional manner, some previous assumptions no longer hold and have to be adapted:

1. No constant transmission timing possible (12)

2. Medium access control

3. Immediate phase jump (22)

4. Dead lock case elimination

**1.**   Messages are not transmitted instantly.  Between the send command on the software layer and the last electromagnetic signal on the radio channel, at least several hundreds of microseconds pass.  Tyrrell et al. propose that it is possible to predict and fix this duration which it takes a network node to transmit a message, i.e. *constant transmission timing*. Maroti et al. (12) have explored internal delays in sensor nodes and found that nonde-terministic delay sources exist, which have to be accounted for.  Figure 3.5 summarizes the decomposition of the delivery delay of an idealized point in a message (e.g. the first bit of a packet) as it traverses through software, hardware and physical layers.  Processor load, and internal interrupts create sources of random delay which limit the precision in synchronization.  While the propagation delay can generally be omitted due to the short distances in WSNs, the medium access delay varies greatly over two orders of magnitude. All delays shown in the figure add to the delay uncertainty, which is present in network nodes.  Therefore, the transmit time **cannot** be assumed to be constant for all messages.

**2.**   Tyrrell et al. suggest to use predefined pseudo noise (PN) sequences which can be recognized by a correlator.  Only a single unique synchronization message should be used in the entire network (independent of message source or destination).  This is not recommended (because it requires special hardware) and prohibits hardware delay compensation.

Differing from Tyrrell et al.'s assumption, MICAz employs a *CSMA* scheme (see Chapter 4.2.3) which introduces random delays of up to 32 ms.  Both of these problems, processing

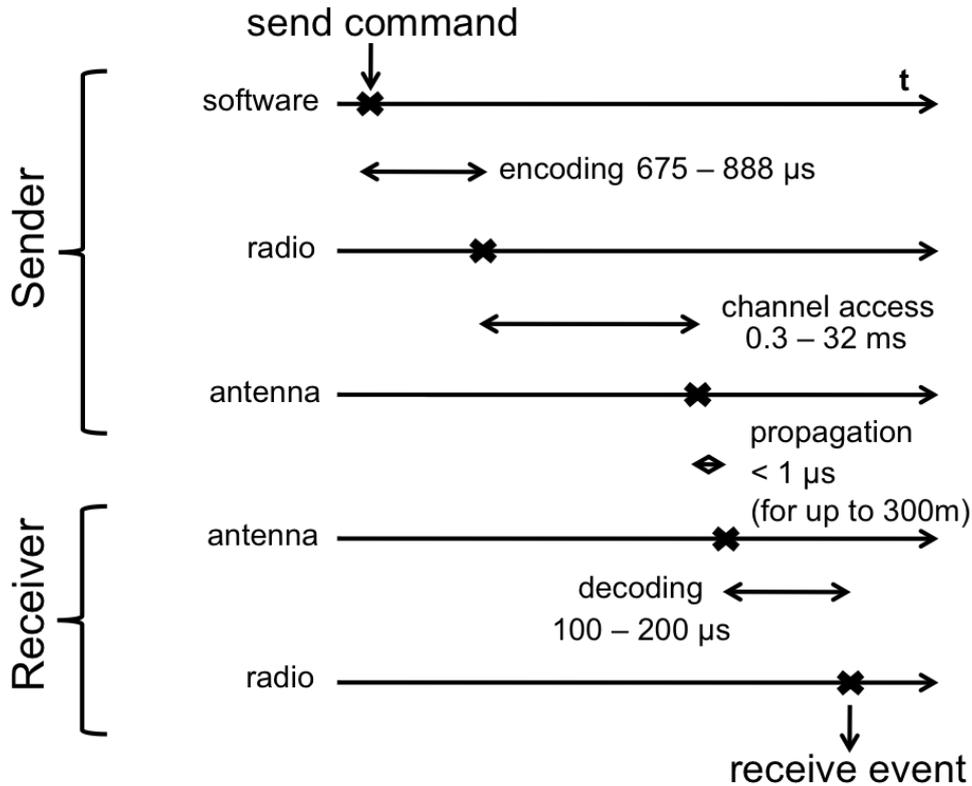

Figure 3.5: Delay estimates for an idealized point of a message (e.g. a data bit) through the components of two network nodes in a transmission system with estimated associated delays

and MAC delay, can be solved by a technique employed in numerous synchronization protocols, called MAC layer time stamping (12; 11; 18), which works as follows:

When a transmit command is issued, the synchronization message is properly established and queued for transmission. However, when - on the physical layer - the transmission of the first bit starts, "last minute" time stamping is performed. In a previously reserved place within the message a time value is inserted while the beginning of the message is already being transmitted. This is done in the memory of the radio chip. The time stamping allows to write additional information into the message, regarding how long the creation and processing of the message took the transmitter. In turn, this enables all receivers of this message to calculate when the send command was issued and eliminates the sender delays, which would otherwise pose the bound on precision.

**3.** Two options exist how a node can react to a received synchronization message. It can either increase its internal phase slightly (suggested by (21)) and reach synchrony over several periods. Or it can immediately adjust its phase to "fire", i.e. perform a *phase* *jump* (22). In reference to the phase diagram, Figure 3.4, a stepwise increase would mean to shorten the *listen* phase slightly after the reception of a message. In case of a phase jump scheme, the *listen* state would be ended immediately, followed by the *wait* state. In this thesis, the immediate phase jump option is chosen to minimize convergence time.

**4.** Tyrrell et al.'s scheme divides the network in two groups. As mentioned earlier, a *dead lock case* exists where all nodes may (by accident) be part of the same group. The probability of this occurrence is highest for scarce networks and depends on the duration of the *refract* state. A solution to this problem is proposed here. The protocol has been adapted to keep a counter of periods which are missing incoming synchronization messages. If for a number of $n_P$ periods, no synchronization message has been received a node may either be out of radio range of other nodes or locked up in the same group, where $n_P$ is a random number between 10 and 25. Therefore, it will repeat its listening state and reset the counter. This assures that after no more than $n_P$ cycles the groups will be separated again.

## 3.6 Firefly 2.0

This section brings the above considerations together, and describes the details of the synchronization protocol as implemented in this thesis. Succinctly, the implementation of the Firefly Protocol on MICAz network nodes required the protocol from Tyrrell et al., the methods regarding delays and precision from Maroti et al. and the infrastructure and control provided by TinyOS. It is called "Firefly 2.0". Only combining these three key elements and the four adaptations from the previous section results in a functioning and precisely running system.

Figure 3.6 goes into detail regarding delay handling as it is performed in Firefly 2.0. The Figure is a more precise view of the transmit state from Figure 3.4. Firstly, a fixed time interval is reserved for the transmission of all synchronization messages, $T_{TX}$. This interval has to be chosen large enough to fit all synchronization messages in a high traffic environment and is determined by the contention of the network. Since a higher node density creates more messages, $T_{TX}$ should be larger for higher densities or the contention backoff must be limited for all nodes. It must under all circumstances be prevented that synchronization messages are still being sent after expiration of the transmit phase.

Whenever a node finishes the *wait* state and enters the *transmit* state, it commands the transmission of a synchronization message and marks the command time with a timestamp $t_A$. After processing (encoding delay) and the initial backoff, the channel will be checked for clearance. Depending on traffic, additional delays may occur through the contention backoff. When the message is then ready to be sent, up to 32 ms may have passed. Now, when the first byte of the message is being transmitted onto the radio channel, it is timestamped for a second time with timestamp $t_B$. The difference of $\Delta t_{TX} = t_B - t_A$ is written into the outgoing message.

In the receiver, a similar procedure is implemented. Upon the first received byte, a timestamp $t_C$ is created. When the message is entirely received, aligned, decoded and found to be a synchronization message, it is recorded again with timestamp $t_D$. $\Delta t_{RX} = t_D - t_C$ is the processing time of the receiver. Knowledge of $\Delta t_{TX}$ and $\Delta t_{RX}$ allows the receiver to infer when the sending command was given. Since the duration of the transmit phase $T_{TX}$ is predefined, the receiver can count down the remainder of this duration and begin the *wait* state at the same instant as the transmitter enters the *refract* state, thus reaching synchrony.

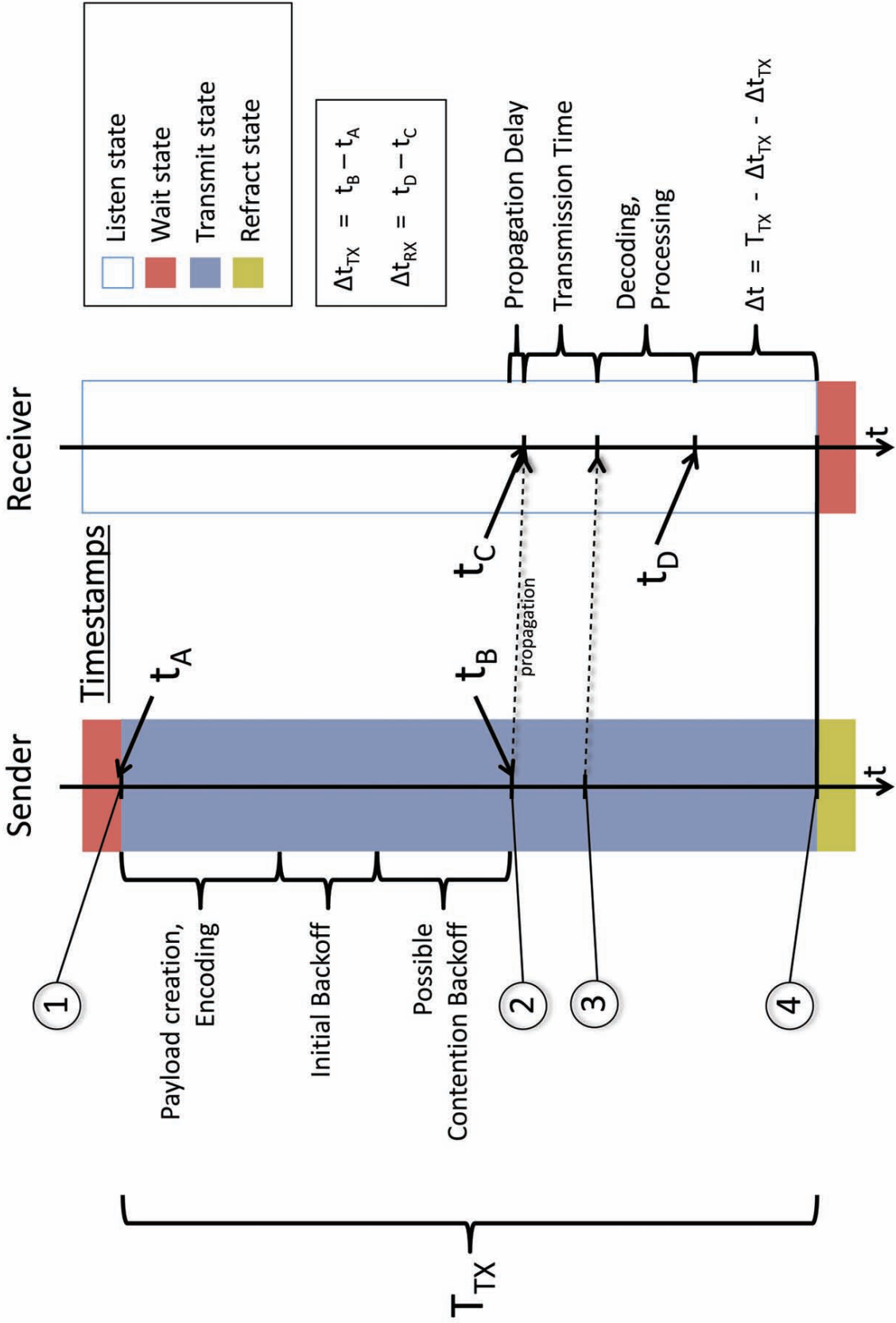

Figure 3.6: Detailed view of the timestamping process in Firefly 2.0. Via four timestamps $t_A$ through $t_D$, the receiver is able to estimate the remainder $\Delta t$. After the expiration of $\Delta t$, both sender and receiver node enter the next phase synchronously. Important points in time:

1 - The Sender's wait state ends and the sending of the synchronization message is initiated.
2 - The first bit of the synchronization message leaves the antenna.
3 - The last bit of the message is transmitted.
4 - After expiration of $\Delta t$ both nodes are synchronized.

# Chapter 4

# The MICAz Network Node

This chapter introduces the MICAz network node (Section 4.1) on which the synchronization protocol for the present work was implemented in detail. It provides all necessary information to perform operations relevant to timing.

## 4.1   TinyOS Programming on MICAz

### IEEE 802.15.4 / ZigBee

"IEEE Std 802.15.4-2003 defined the protocol and compatible interconnection for data communication devices using low-data-rate, low-power, and low-complexity short-range radio frequency (RF) transmissions in a wireless personal area network (WPAN)."[1] The IEEE standard specifies the network topologies, architecture, MAC aspects as well as physical layer details necessary to design a compliant network node. All Berkeley motes (like MICAz) are designed according to this standard.

The IEEE standard is extended by the ZigBee specification[2] which enables network extensions like discovery, multicasting or security. It is not directly involved in this synchronization protocol and is mentioned here only for a complete description of the MICAz network node.

### MICAz

The MICAz (Figure 4.1) is the latest generation of MICA Motes from Crossbow Technology[3]. The term "mote" refers to a node in a wireless sensor network and was coined during research at the University of California, Berkeley, in the 1990s. The term "Berkeley mote" is often used when referring to a network node with similar capabilities independent of the product family.

---

[1]http://ieeexplore.ieee.org/servlet/opac?punumber=4299494

[2]www.zigbee.org

[3]www.xbow.com

| Parameter | Value |
|---|---|
| Processor | Atmel ATMega 128L, 8 MHz |
| Program flash memory | 128 kB |
| Serial flash memory | 512 kB |
| EEPROM | 4 kB |
| Analog to Digital Converter (ADC) | 10 bit |
| Radio Chip | Chipcon CC2420 |
| Radio Frequency | 2.4 GHz to 2.4835 GHz |
| Data Rate | 250 kbit/s |
| Power Consumption | 19.7 mA (receive) |
|  | 17.4 mA (transmit, 0 dBm) |
|  | $< 2 \mu$A (sleep) |

Table 4.1: MICAz hardware parameters

The MICAz specifics attempt to combine the requirements of applications (computation power), battery lifetime (minimal energy consumption) and researchers (flexibility, handling). Therefore, the computing capacity is a tradeoff between saving energy while allowing a certain complexity in applications and data storage. The platform version for developers is programmable through a USB board. For hardware parameters, see Table 4.1[4].

The mote transmits at a frequency of 2.4 GHz with a raw data rate of 250 kbit/s using a carrier sense multiple access collision avoidance (CSMA/CA) scheme. Three light emitting diodes (LED) allow limited user interaction. The transmission power (and thus the range) is variable to preserve energy and adjust the system setup to the laboratory.

**TinyOS**

Programming of the protocol was done using the Tiny Operating System (TinyOS), an open-source OS specifically designed for networked embedded sensors[5]. It handles the severe memory and power constraints with event-centric concurrent applications and a truly tiny OS core of only 400 bytes. For this reason, it has become the favored operating system in WSNs. In November 2006, TinyOS version 2.0 has been released which is used for this work.

---

[4]MICAz Datasheet:
http://www.xbow.com/Products/Product_pdf_files/Wireless_pdf/MICAz_Datasheet.pdf
[5]www.tinyos.net

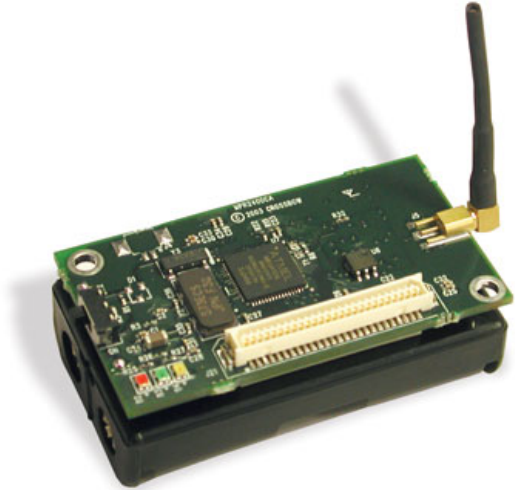

Figure 4.1: The MICAz network node (Source: www.xbow.com)

In order to minimize the code size on a mote, TinyOS is not an operating system in the traditional sense which is entirely installed on a mote. It is rather a framework providing building blocks for specific applications. At compile time only the specified components will be integrated into an application, therefore keeping the binary memory requirements at an absolute minimum.

TinyOS programs are written in a C dialect called "nesC"[6] (nesC = networked systems C). The important innovations for this work are the separation of construction and composition and interfaces. In nesC components can be "wired" together in the application file, thus allowing to use previously created modules by simply "wiring" them together. Specific additions to the code are then made in the construction file. Interfaces allow to interact with "black box" components without having to go into their source code. They enable to make a *call* to a command or receive an *event*. Often, interfaces do not provide the desired functionality and have to be extended manually. Since TinyOS is open source, this is possible, yet time consuming, as components can be quite vast.

## 4.2 Implementation Issues with Synchronization

### 4.2.1 Simulation of the Model

The first step to evaluate a theoretical model is simulation. While the original authors of the Firefly Protocol have simulated it in MATLAB, this does not reflect hardware effects like delays, which are the central issue of this thesis. Another option, here, is TOSSIM (33), a simulator provided along with TinyOS (TOSSIM = TinyOS SIMulator). It permits increased interaction capabilities which is especially useful for debugging, as it is particularly difficult to find the source of an error in a network node, if the only human interaction interfaces are three LEDs. However, TOSSIM does not implement time consumption and therefore assumes every function call to finish instantly. Also, physical network aspects like medium access control are only marginally fitted. For these reasons it was not used in this work and is only mentioned for completeness.

---

[6]nescc.sourceforge.net

## 4.2.2 Time and Clocking on MICAz

Regarding timing, no interfaces or components are provided by TinyOS that go beyond the absolute basics. There is no internal representation for time stamps like minutes, hours or days. Probably, because of the numerous platforms which TinyOS supports and the requirements towards precision that are not yet defined, this functionality is still to be supplied.

All clocking and timing functions on a MICAz radio mote are provided by the onboard microcontroller, ATmega128L[7]. The microcontroller is clocked at $f_0 = 7.3728$ MHz. All timed operations base on $f_0$, but different derivations exist. The flexible millisecond timer (named TimerMilliC in TinyOS) increments every 32 ticks of a Jiffy = 32,768 Hz = 30.518 $\mu$s, i.e. 1024 times per second. The more precise microsecond counter (CounterMicro32C) runs at $f_1 = f_0/8 = 0.9216$ Mhz (Table 4.2). This counter is linked to the chip over compare registers. Therefore, the number of simultaneous counters is limited to three, whereas a virtually unlimited number of timers can be created. After initially using both timers in this project, the program was later slimmed to run all operations on a single microsecond counter.

| Name | MilliC | 32kHz | MicroC |
|------|--------|-------|--------|
| Frequency | 1,024 Hz | 32,768 Hz | 0.9216 MHz |
| Period | 976 ms | 30.518 $\mu$s | 1.084 $\mu$s |

Table 4.2: Available timer precisions in TinyOS/MICAz

This leads to the fact that all clock values used in this thesis are of unit **clock ticks**, which is

$$1 \text{ clocktick} = 1/f_1 = 1/0.9216 \text{ MHz} = 1.085 \mu s.$$

The timer and counter components are precise up to the accuracy of the system clock. The initial starting of a timer costs between 100 and 500 $\mu$s. However, every successive start or stop of that timer is correct to the microsecond.

---

[7]http://www.atmel.com/dyn/resources/prod_documents/doc2467.pdf

**Estimating Clock Drift** In the laboratory, a setup was created in which all nodes in the network report their perceived duration of a reference (see Figure 4.2). The base station transmitted two beacon messages (1 second apart) and waits 10 seconds before sending the next beacon pair. (These durations were chosen to minimize collision packet loss while still allowing compact test periods.) All slave nodes receive the beacons, time stamp their arrival times $t_{R1}$ and $t_{R2}$ and report the perceived difference $\Delta t = t_{R2} - t_{R2}$ back to the base station. Due to the inherent clock drifts in each node, the reported durations differ slightly. Table 4.3 shows the resulting offsets. Within the one second interval, the clocks of the reference and the nodes on average differed by up to 13 clock ticks.

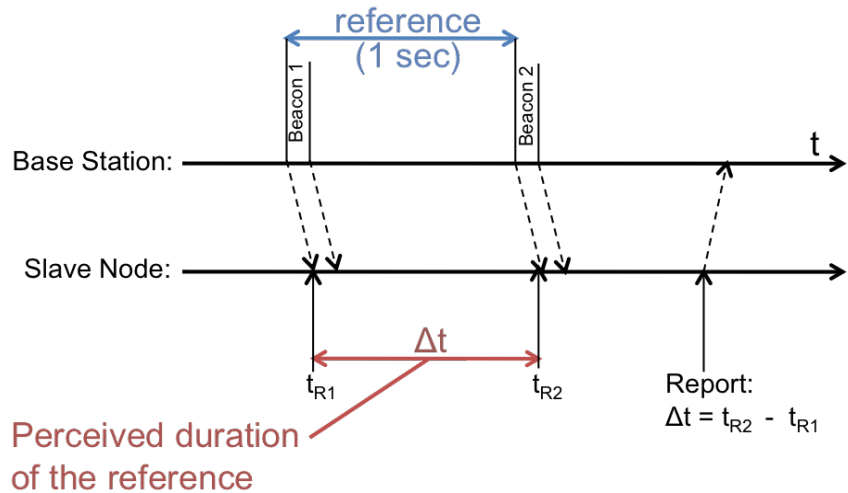

Figure 4.2: Measurement setup for perceived timings in network nodes

| Node ID | 1 | 2 | 3 | 4 | 5 |
|---|---|---|---|---|---|
| Clock drift per second [clock ticks] | 8 | 7 | 13 | 6 | 5 |

Table 4.3: Results of the internal clock comparison in MICAz motes

## 4.2.3 Medium Access Delay

In a system which bases on the concept of periodic and precise flashes of light it makes a large difference that the radio channel has to be shared. Whereas light flashes (generally) do not interfere or eliminate each other, radio messages will do so. Two non-identical radio messages which overlap on the same channel will cancel each other out. Therefore, the radio channel has to be shared and controlled by a Medium Access Control (MAC) layer. In the case of MICAz the MAC system is Carrier Sense Multiple Access (CSMA).

CSMA in the case of MICAz is simplified as compared to more advanced systems. When a packet is enqueued to be transmitted, it will first receive a uniformly distributed random time offset (Initial Backoff) with limits defined in the IEEE 802.15.4 standard. After this initial delay, the transmitter checks the channel for transmissions. In case the Clear Channel Assessment (CCA) returns a busy channel, the packet will once again be put on hold for a random amount of time (Congestion Backoff). This procedure is repeated until the channel is found to be clear and the packet can be transmitted. The risk of two motes both simultaneously recognizing the channel as free and transmitting - thus eliminating each other - still exists. Also, other MAC specific problems like the *Hidden Terminal* or message acknowledgment are not addressed in the standard and are left to the application layer (34; 35).

In MICAz motes, unexpectedly, the CSMA protocol is software controlled. This means that the radio chip takes orders from the CPU and reports back with flags and will not make decisions on its own. The advantage is here, that the programmer can take direct influence on the CCA and backoff intervals over the RadioBackoff interface. For each individual message, CCA can be allowed or disallowed and the backoffs separately configured. This information is contained within the TinyOS system libraries for the CC2420 chipset.

**Estimating MAC delay**  A MICAz mote was configured to periodically send radio messages. The time difference between the send command and the actual physical sending of the first byte was recorded and reported to a PC for offline evaluation. Figure 4.3(a) shows the histogram of the measured transmission delays without traffic. The mote was shielded to prevent interfering signals from Wireless LAN (IEEE 802.11) which operates on the same frequency.

Next, a traffic scenario was created. To maximize the contention backoff, five motes were programmed to continually send back-to-back messages on the same channel. These traffic sources forced the base station mote to use very high contention backoffs. The resulting histogram can be seen in Figure 4.3(b)

Following the IEEE 802.15.4 standard, the default for the initial backoff (IB) is a uniform distribution between 10 - 320 *Jiffies* and for the contention backoff (CB) is a uniform distribution between 10 - 80 *Jiffies*. With 1 *Jiffy* = 30.518 $\mu s$:

$$IB = [305 \ \mu s; 9766 \ \mu s] \quad \text{and} \quad CB = [305 \ \mu s; 2441 \ \mu s].$$

Since CSMA introduces significant random message delays into the network it has to be compensated for in the protocol.

The option of disabling CSMA selectively for synchronization messages was studied. It was found that by turning off the CCA for particular packets, the total sender delay could be reduced to a constant $655 \mu s$. However, the resulting packet loss would render the synchronization system useless. Therefore, the medium access delays are compensated through precise timing and timestamping.

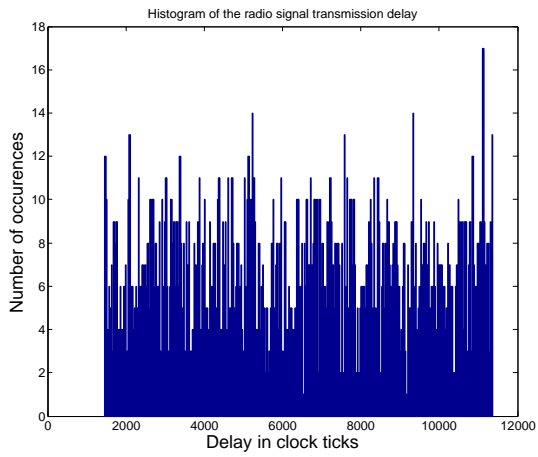 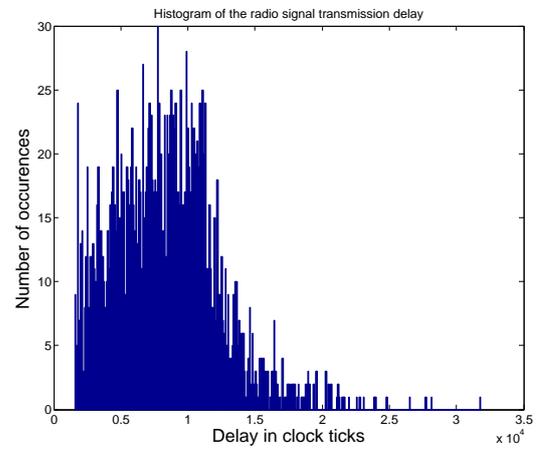

(a) Transmission histogram of a MICAz mote in a low traffic scenario (i.e. no contention backoff.

(b) Transmission histogram of a MICAz mote in a high traffic scenario. Five additional motes continually resend messages on the channel. Delays of up to 32 ms occur.

Figure 4.3: The peaks are results of the pseudo-randomness of the number generator and the bucketing procedure of the histogram. The initial processing delay of the hardware can clearly be seen up to 1450 clock ticks.

# Chapter 5

# Implementation

Up to this point, the necessary background for the implementation has been provided. This chapter covers what implementation details were necessary to realize Firefly Synchronization on MICAz motes, and summarizes the central issues from chapter 4.

## 5.1  Important Issues

**Communication**  Firstly, the behavior of MICAz motes regarding communication is not distinct, since the IEEE standard 802.15.4 leaves many options open which can be configured on the software level. It was found that acknowledgment of messages is not inherently active and the nodes follow a "fire and forget" scheme. Therefore, packet loss can not be detected by a node. For synchronization, this is advantageous since a possible re-request of messages introduces large delays which need to be avoided if possible. Also, there is no limit on the number of contention backoffs taken when the channel is found to be busy. Theoretically, if the channel were continuously busy, a node would never send a packet and at the same time never drop the packet.

When running unmodified, frequent packet losses occur because of the simple CSMA scheme applied. Therefore, to increase packet delivery reliability it is useful to add an additional random component to transmit times even before the CSMA process is initiated. In addition to the radio channel packet loss, the serial port poses a bottleneck. It has a lower bit-rate than the radio channel and in the case of burst traffic (at the synchronization points) frequent packet losses occur. This problem was solved by adding a packet queue to the base station mote.

**TinyOS version 2.x**  Secondly, major changes were made to the system core of TinyOS during the transition from version 1.x to 2.x. Source code for version 1.x will not compile under 2.x. This means that any applications developed before the release of 2.x have to be converted and to date only few applications exist for 2.x. Many issues once solved have to be resolved in new data structures. For example, access to the MAC layer time stamping was changed significantly.

**Time representation**  When implementing the protocol on the wireless nodes, it was unclear how the representation of time works inside of a MICAz mote. No specific interface exists which allows for seamless access, conversion and modification of time. Three precisions of counting exist, milliseconds, 32 kHz (i.e. 32,768 clock ticks per second), and microseconds. These options were explored and the decision was clearly made for microsecond precision which offers the best resolution. For this accuracy, the counter component (which counts up and can be read out) and the alarm component (which counts down and fires on reaching zero) can be used.

Next, several components were explored which had to be discarded because of limitations. For example, the TinyOS provides a RadioTimeStamping interface, which allows to timestamp incoming and outgoing radio messages. However, this timestamp is of size 8 bit = 65,536 values. When running at 32 kHz, this creates an overflow every 2 seconds. Since synchronization intervals are much longer than this, this interface is not useful and the timestamping was implemented manually.

**Disabling the medium access control**  As described in Section 4.2.3 it is important to estimate medium access which creates the dominating delays. An attempt was made to follow Tyrrell et al.'s suggestions (19) of fixed transmit times by disabling the clear channel assessment (CCA) of the CSMA scheme. This is possible, as the entire scheme is implemented on software level. However, packet losses due to collisions by far outweighed possible gains of the CCA shut-off. Therefore, timestamping was used instead.

**MAC layer timestamping**  The previously mentioned and critically important *MAC layer timestamping* is possible via direct access to the radio chip (CC2420). The chip designers offer an interface, the *CC2420Transmit*, which allows to access the send buffer from software. If the internal bit offset is known, pieces of the send buffer memory can be replaced. In this particular case it works as follows: An IEEE 802.15.4 radio message consists of a header and a data field. The header contains 11 bytes of meta information like, for example, message length, message source or message type. The data field is of variable length. Knowledge of the header size and the position of the desired data piece inside the message allow to modify a particular group of bytes in the memory. Extreme care has to be taken regarding this offset and the endianness of the layers in between. TinyOS provides the *network data type* which internally corrects the endianness as necessary.

Practically, the operation runs as follows: On the software level a send command is issued. The data payload is constructed, filled and added to the header. The message is passed to the radio chip. The radio chip initiates the channel assessment of the CSMA scheme.

When the channel is found to be clear, the software layer is informed and calls the memory modification function. This modification is, therefore, performed in the last moment when the header of the message is already being transmitted.

**Files** The protocol was programmed as a compact nesC component consisting of a header file, a configuration file and an implementation file[1]. The transition through the phase states was realized through an alarm with microsecond precision that was instructed to start, stop and reset as needed upon state changes. All nodes are programmed identically.

## 5.2 Measurement Setup

In order to evaluate the performance of the Firefly network, the synchronization data and precision need to be collected by a central entity. This entity is a Unix operated PC which connects to a passive base station mote via the USB port. The base station mote collects all radio messages in range, timestamps their arrival and forwards them to the PC for offline evaluation. The constructed files are then interpreted in MATLAB 7.0. (See Figure 5.1).

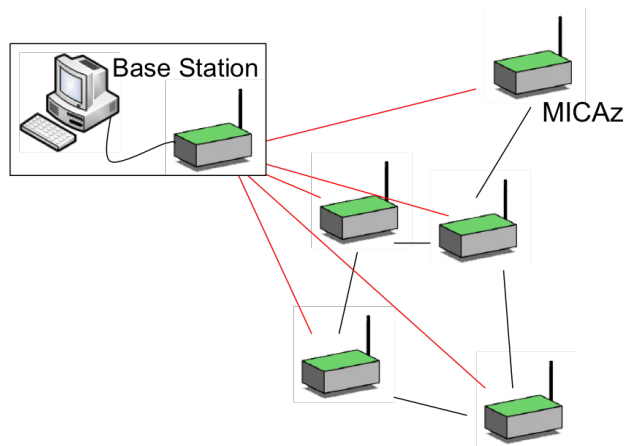

Figure 5.1: Laboratory setup for the evaluation of the Firefly Protocol (simplified)

---

[1]All files are provided on the attached CD-ROM.

# Chapter 6

# Results

Because of the inherent robustness and flexibility within the Firefly network, it is extremely difficult to evaluate. Since there exists no external or network wide reference, what point in time should be used for comparison? And since nodes are free to make mistakes or switch groups, there is no immediate right or wrong for a mote, as long as the network as a whole is clocked correctly. Since this approach creates a dynamical system, it is also very difficult to explain and predict its behavior. Small offsets or delays of a single node may disturb the entire system for several cycles.

To categorize and evaluate systems with a different number of nodes, the following criteria and plots will be discussed:

1. **Mean and variance** of the offsets from the reference

2. **Probability Distribution Function** (PDF) of the offsets on a millisecond and microsecond scale, compared with the offset positions in each node

3. **Cycle durations** over time

4. **Offsets** from the reference plotted **over time** (to recognize specific network behavior)

Before going into detailed scenarios, the term "reference" has to be clarified as it is not given by the system and has to be defined.

**Regarding the Reference** The point of reference in this evaluation is the average of the "firing points", in between 600 ms intervals. This means that if the base station (as data collector) does not receive any messages for at least 600 ms, it considers it the start of a new interval. Within this interval, all message timestamps are averaged. This average is treated as the reference from which the time offsets are calculated. For example, for three nodes firing at $t_i = \{2.9, 3.0, 3.1\}$ the reference becomes $t_{ref} = 3$ and the offsets are $o_i = \{-0.1, 0, 0.1\}$. Since in this evaluation the average offset is equal to zero due to symmetry, all offset statistics are created from absolute values. In the example, the average offset is $\overline{\|o_i\|} = 0.1$.

Since the base station is a passive reader of the synchronization messages, it must be in range of all network nodes. Thus, all systems analyzed are single hop networks with all nodes in range to each other. Such networks are also called fully meshed networks in literature.

Cycle durations are calculated as the differences between the reference points in each group. Thus, a "cycle" is a complete tour around the phase state circle (Figure 3.4) - not the half circle between two adjacent firings of Group A and Group B. A change in cycle durations can shed light on a constant drift of the network due to, for example, asymmetries in the synchronization period.

As stated in Section 4.2.2, *clock ticks* are the basis for all timing operations in a node and have a duration of 1.085 $\mu s$. All measurements and operations base on this unit. However, for the sake of simplicity and without loss of generality, clock ticks will in this section be treated as having a duration of 1 $\mu s$. This makes plots and scales easier to understand. Since all nodes and instruments in the system count similarly, all results scale correctly to each other and differ from reality through the factor 1.085.

The analysis is performed stepwise - with a higher number of nodes in each experiment - to go into specific aspects in turn as they appear in the more complicated systems.

## 6.1 Single Node Evaluation

First, the behavior of a single isolated node is shown in Figure 6.1 to visualize the counter-dead lock mechanism. When started, the node runs with a period of 2,000,330 $\mu s$. (With $T = 1\ s$ the node requires $2\ T$ to go once around the phase cycle.) It is important to note that although all internal timers are set to have $T = 1,000,000\ \mu s$, processing of code instructions already increases the idle run time of a node by 330 $\mu s$ (not shown). However, this delay is deterministic and remains constant over all cycles. The idle cycle duration is precise up to $\pm 1\ \mu s$, which is the limit of the internal clock. Since the node is isolated, it does not receive any synchronization messages. In order to prevent a dead lock case (see Section 3.5), it remains in the listen state for an extra second, thus increasing the period to 3 seconds occasionally.

Offset evaluation is, of course, not possible in the single node case.

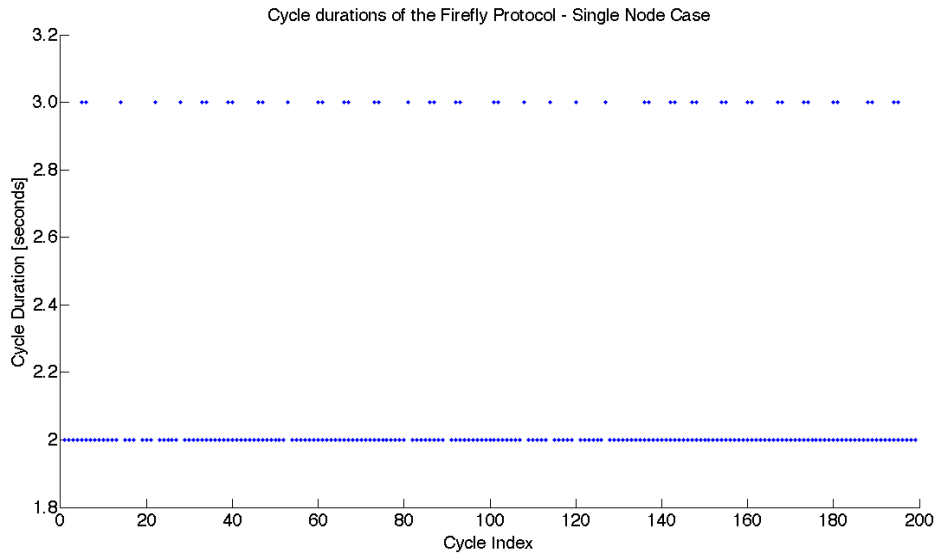

Figure 6.1: Cycle duration for a single isolated node. In most cases, the node ran with a period of approximately 2 seconds. It can be clearly seen that occasionally the node repeats half of a period, amounting to 3 seconds. This is caused by the node trying to break a deadlock case as described in Section 3.5

## 6.2   Dual Node Evaluation

Next, a scenario with two nodes is recorded for 200 cycles which gives a first impression of cycle durations in a dynamic system. The nodes are intentionally booted at the same time, thus creating a dead lock state. It takes 11 cycles until the first node switches groups and resolves the dead lock. In cycle 82 another dead lock occurs, which is resolved in cycle 90. Figure 6.2(a) shows the duration of the cycles. Although some cycles appear to have a 4 s period, this is an effect caused by the base station. Since the base station expects two groups (1 s apart), but only sees one group with two members, it (falsely) assumes that there are two groups with two members. Since this plot compares the cycle durations within each group, they appear to be 4 s long during dead lock states. Although not intended, this glitch makes it very easy to recognize a deadlock state in the cycle duration plot. As soon as the dead lock is resolved, the base station marks the groups correctly and cycle durations appear to be back to 2 s (again, not precisely 2 s, but close enough for the dead lock considerations).

A closer look at the cycle durations which are not dead locks or outliers reveals that the cycles are precise up to $\pm 1$ $\mu s$ around $2,000,078$ $\mu s$ (Figure 6.2(b)) and remain stable over time. Occasionally, the system makes mistakes leading to wrong cycle durations.

Since in the dual node system, each group has only one member, each node is its own reference and has no offset. Therefore, offset evaluations are not yet possible.

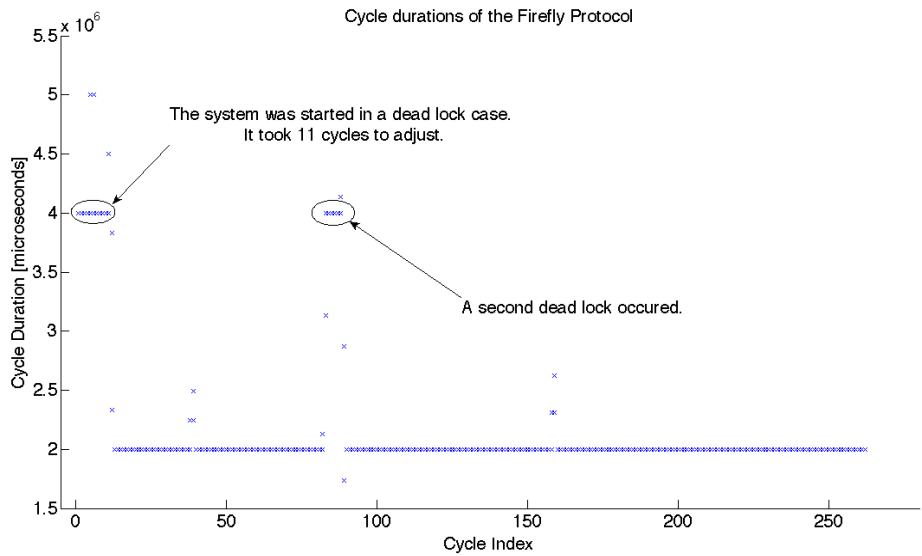

(a) The cycle durations illustrate how the systems behaves over time. The cycle duration during the dead lock is not actually twice as long, but is specifically marked by the base station to allow for an easier display.

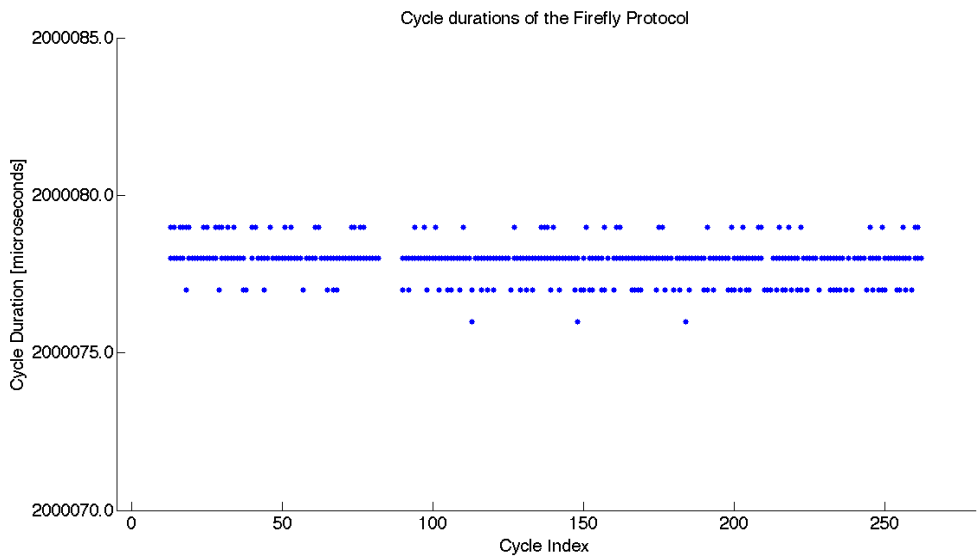

(b) On the small scale, cycle durations are precise within $\pm 1$ $\mu s$. The cycle durations are 78 $\mu s$ longer than the designated 2 s due to small software processing delays (similar to the idle cycle duration).

Figure 6.2: Evaluation of the two node Firefly system

## 6.3   Four Node System Evaluation

The four node system is analyzed, because it reaches maximum precision. The system is large enough to have a group size > 1 , yet small enough to have a low probability of random outliers. Figure 6.3 shows that on the millisecond scale, outliers still go as far as 600 ms. However, 99 % of the values lie within 100 $\mu s$ of each other, and 95 % are within 5 $\mu s$. Disregarding outliers with an offset of more than 5 ms, this results in average offsets between 1.5 $\mu s$ and 4.5 $\mu s$ (see Table 6.1).

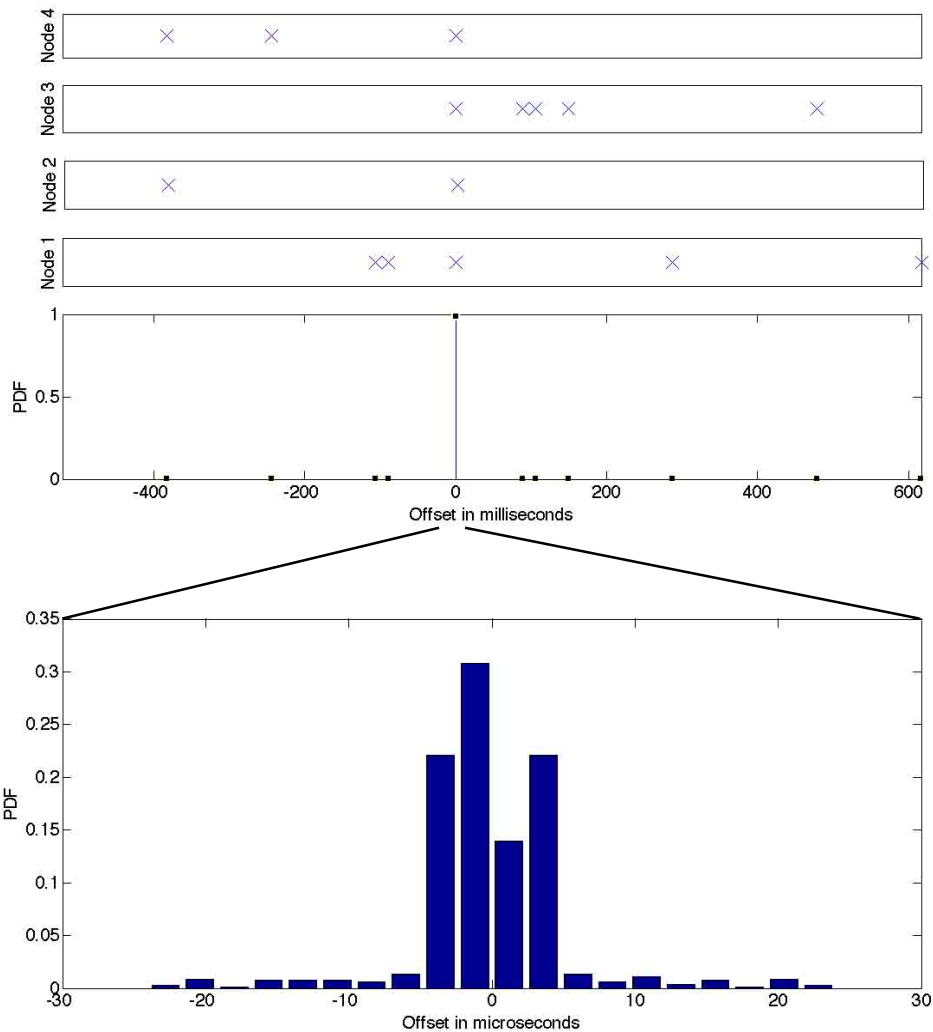

Figure 6.3: Offset values and distribution of the four node Firefly system

Regarding cycle durations, details can be taken from Figure 6.4. Since the cycle durations are calculated as the difference of two reference points, they are not constant and vary within 20 - 30 $\mu s$ of the mean at $2,000,084$ $\mu s$. There are two dominating cycle durations at $2,000,088$ $\mu s$ and $2,000,080$ $\mu s$. This is explained in the following: From Figure 6.5 it can be seen that the nodes Group B constantly fire 8 $\mu s$ apart from each other. Now, Group A will adjust its phase according to a synchronization message it receives from Group B. Due to the random initial backoff during channel access it is random which node fires its synchronization message first. Therefore, Group A will sometimes use the early message from node 2 and sometimes the late message from node 4 as the basis for the phase adjustment. Therefore, this error of 8 $\mu s$ is carried into the other group, thus creating two equally probable cycle durations. The cycle durations are not related to the groups. That means both groups have equal probability for all cycle durations. In comparison, in the long run there is no evident drift.

|       | Mean $\mu$ of Offsets from Reference | Standard Deviation $\sigma$ of Offsets from Reference |
|-------|-------------------------------------|------------------------------------------------------|
| Node 1 | 1.50 $\mu s$ | 4.37 $\mu s$ |
| Node 2 | 4.38 $\mu s$ | 1.97 $\mu s$ |
| Node 3 | 1.59 $\mu s$ | 4.42 $\mu s$ |
| Node 4 | 4.50 $\mu s$ | 1.73 $\mu s$ |

Table 6.1: Statistics of the four node Firefly system (excluding outliers > 5 ms)

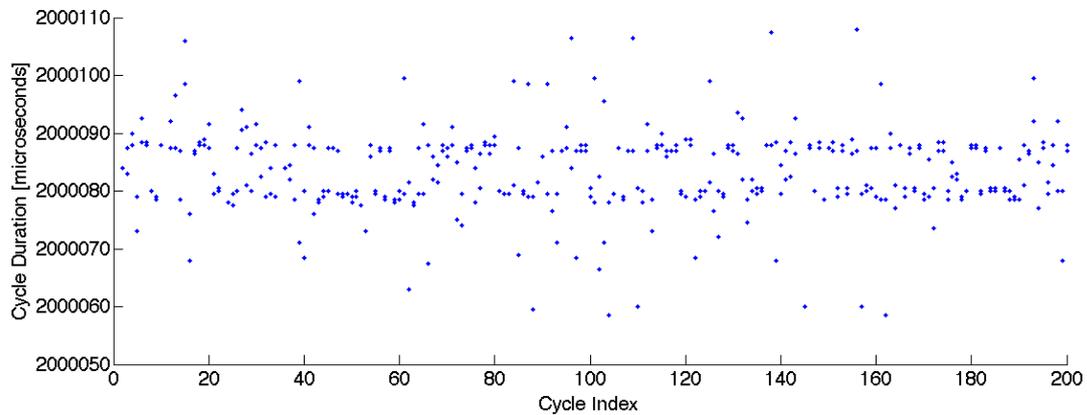

Figure 6.4: Cycle durations of the four node Firefly system. Two equally probable durations stand out.

In Figure 6.5, the offset behavior over time is plotted. It can be clearly seen, that the offsets are symmetric in regard to the reference for each group, due to the calculation of the reference as the average. It has be kept in mind that the reason for this phenomenon is the data processing and evaluation, not the system itself. The omnipresent difference of a couple of microseconds between each node's offsets is accounted to the natural clock drift. Since each node perceives time progression slightly different, this error persists over all cycles.

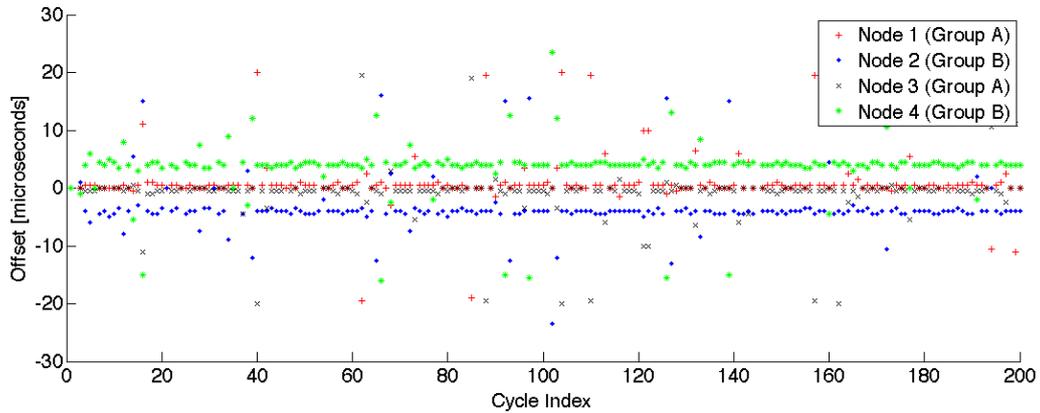

Figure 6.5: Offsets in the four node Firefly system

## 6.4 Five Node System Evaluation

The five node system is chosen to inspect a more chaotic network and shows that the dynamics of the system quickly grow and make it harder to explain what happens in detail. In Figure 6.6 the offset distribution is shown to visualize probabilities. In 60 % of the cycles over all nodes, the offset was within one millisecond of the reference with 40 % being within 5 $\mu s$. This is reflected in Figure 6.7, which displays the behavior of firing offsets from the reference over time. It can be seen that nodes 2 and 5 which were in the same group are extremely precise around the reference which amounts to 40 % of the measurement points. Nodes 1, 3 and 4 from the opposite group, however, have a larger offset.

A noteworthy phenomenon is the belt-like empty space around the reference. It has the proportions of approximately 500 $\mu s$ above, and 1000 $\mu s$ below the reference. This ratio of 1:2 is caused by the fact that always two nodes from Group A have the same (late) offset, while the third node from Group A has a timestamp that is roughly 2 ms earlier. The calculation of the reference as the average, thus, creates the belt-like formation with a tendency to positive (late) offsets. In summary, this means that Group B is reliably precise, but, unexpectedly, does not pass this precision to Group A. From Group A, two nodes are always late in comparison, while one node fires about 2 ms earlier. It is at this point unclear where the technical cause for this phenomenon lies.

The recording started before the nodes were switched on to include the convergence synchrony. However, as predicted, the convergence time is optimally short (i.e. one cycle) and no "tuning" can be seen (i.e. the offset behavior does not improve over time). After converging, the network is reliably within 2.5 ms of the reference and does not evolve. Also, no dead locks occur, since they become more unlikely with increasing node density. Table 6.2 summarizes the statistics of the 5-node network, disregarding outliers beyond 5 ms.

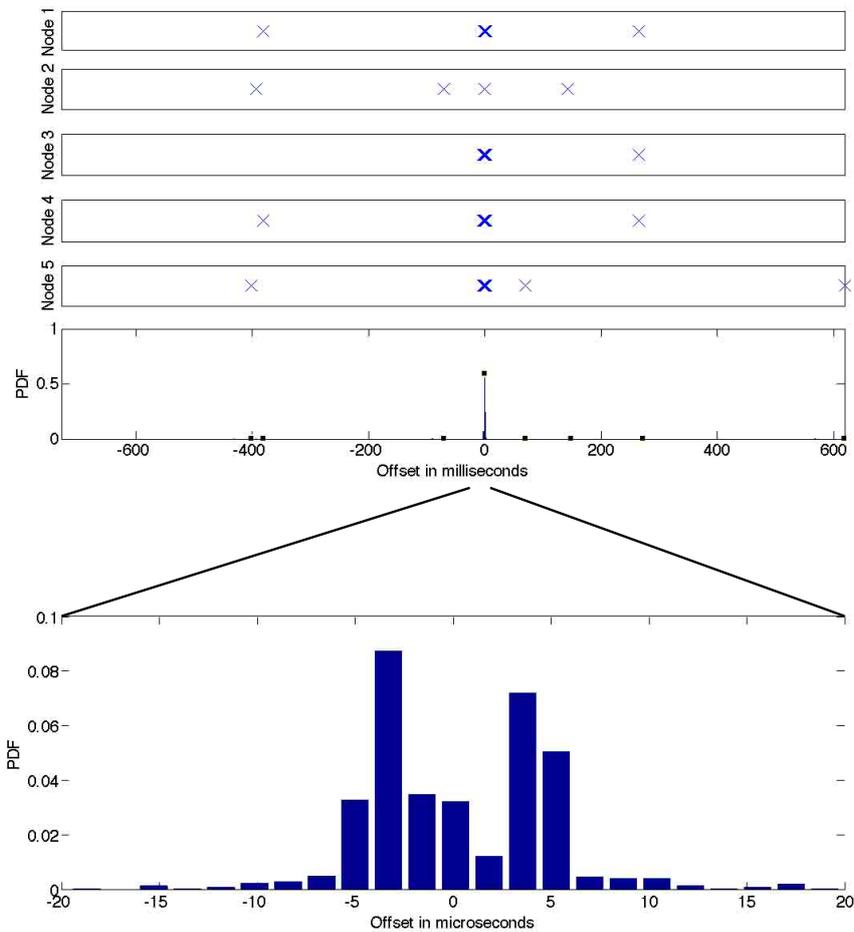

Figure 6.6: Recorded offsets from the reference in milliseconds and probability distribution in a five node system. In 60 % of the cases, the offset was within 1 ms of the reference, in 40 % within 5 $\mu s$.

The cycle durations in the five node network remain within 15 $\mu s$ around the mean at $2,000,084$ $\mu s$, which is the same mean as in the four node system. The characteristic of two similarly probable cycle durations has disappeared. No drifts can be recognized in the long run which leads to the conclusion that the system keeps itself stable through the continuous updates and does not succumb to offsets in either direction.

## 6.5 28 Node System Evaluation

Lastly, an experiment with 28 Firefly nodes was run. The first observation that was made, was the large proportion of packets which did not reach the data sink file. With 28 nodes running for 467 cycles, there should have been just over 13,000 synchronization messages. However, only 5881 were written into the data file - yielding a drop rate of 55 %. (In comparison, in the five node system, 95 % of the packets were received.) This could have several causes. Although the base station keeps a packet queue to relieve the bottleneck serial port, packets may get dropped in the serial queue. Supporting this assumption, the Java application on the PC side gave "bad packet" reports. Another source of error is the channel access scheme. With 14 nodes (if the group membership is assumed to be 50 %) trying to send a message at almost the same instant, the probability of a collision after a positive CCA is evidently increased. However, since the message loss is random due to the random initial backoff of the MAC layer, the general picture is not affected. Therefore, the cause of the packet loss is not explored here.

|          | Mean $\mu$ of Offsets from Reference | Standard Deviation $\sigma$ of Offsets from Reference |
|----------|------------------------|-------------------------------|
| Node 1   | 752.54 $\mu s$         | 609.60 $\mu s$                |
| Node 2   | 3.75 $\mu s$           | 2.00 $\mu s$                  |
| Node 3   | 756.02 $\mu s$         | 761.12 $\mu s$                |
| Node 4   | 885.04 $\mu s$         | 734.85 $\mu s$                |
| Node 5   | 49.25 $\mu s$          | 0.18 $\mu s$                  |

Table 6.2: Statistics of the five node Firefly system (disregarding outliers > 5 ms)

With increasing node number, the precision is dropping. Over all nodes and cycles, the average offset from the reference was found to be around 2 ms (standard deviation approx. 1.5 ms). Figure 6.9 illustrates that most of the offsets were within 10 ms of the reference. This is confirmed by Figure 6.10 which shows the behavior of offsets over 450 cycles. Due to the large number of nodes, only the two groups are marked specifically (as compared to each node separately). This emphasis allows to immediately see that the far outliers always happen to several nodes of the same group at the same time (blue and red groups of outliers, respectively). However, it cannot be concluded that errors propagate through the groups in the sense that an outlier in Group A will cause an outlier in Group B in the next cycle. The outliers appear to be unrelated.

Cycle durations remain within 2 ms of the reference (Figure 6.11(b)), seemingly uncorrelated. It can be seen on the large scale (Figure 6.11(a)) that outliers occur more often than in smaller networks and that the cycles have a tendency to rush, thus having a duration shorter than 2 s. Both figures have the cycle durations of each group plotted in a different color.

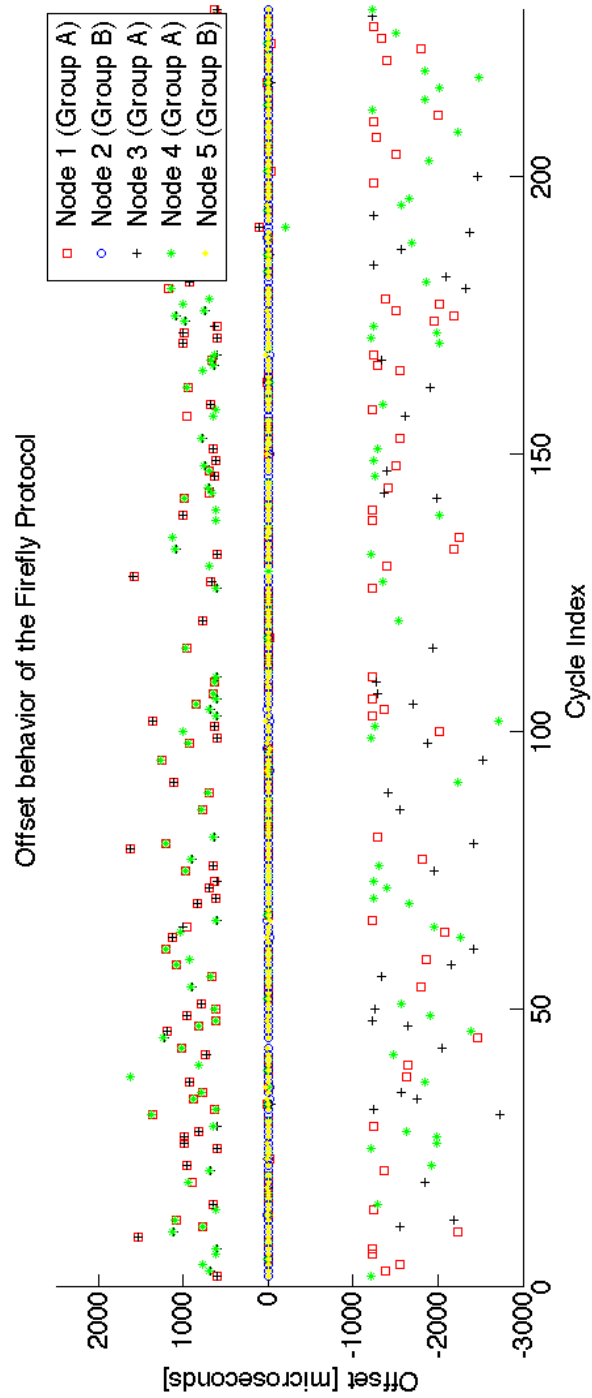

Figure 6.7:  Evolution of offsets over time in a five node system.  Notably, Group B retains maximum precision while Group A makes systematic errors.

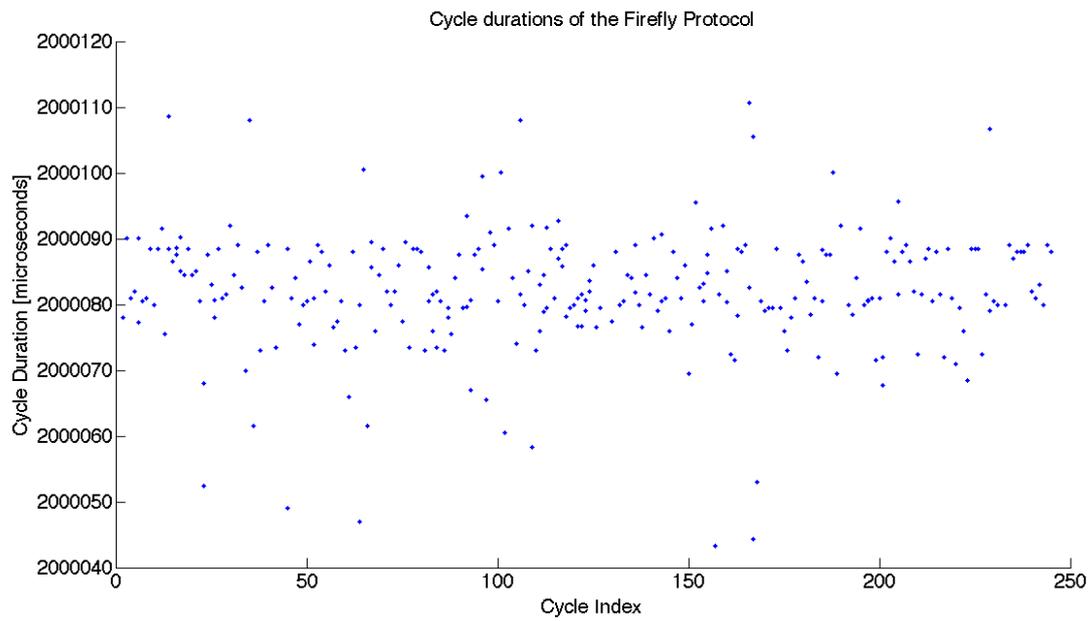

Figure 6.8: Evolution of cycle duration over time in a five node system

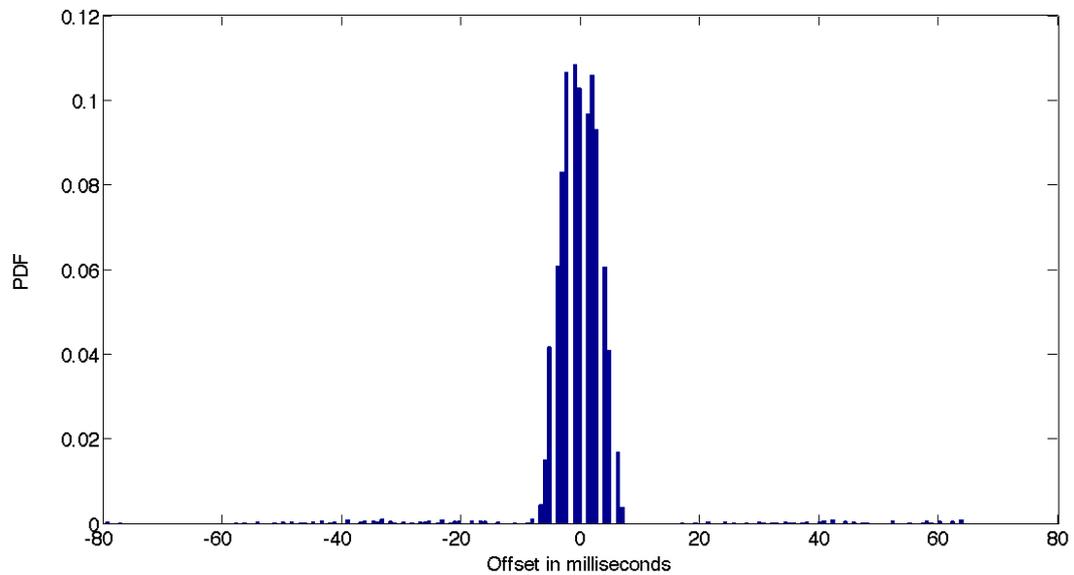

Figure 6.9: Probability density function of the offsets in the 28 node Firefly system

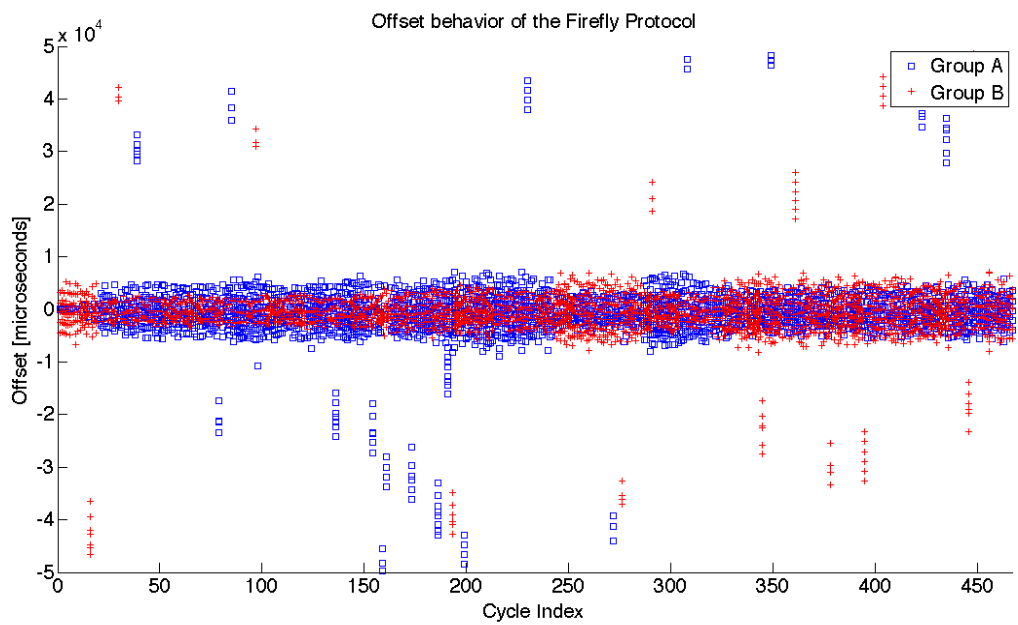

Figure 6.10: Offsets over time in the 28 node Firefly system marked by group membership. Timing errors always happen to several members of the same group at once.

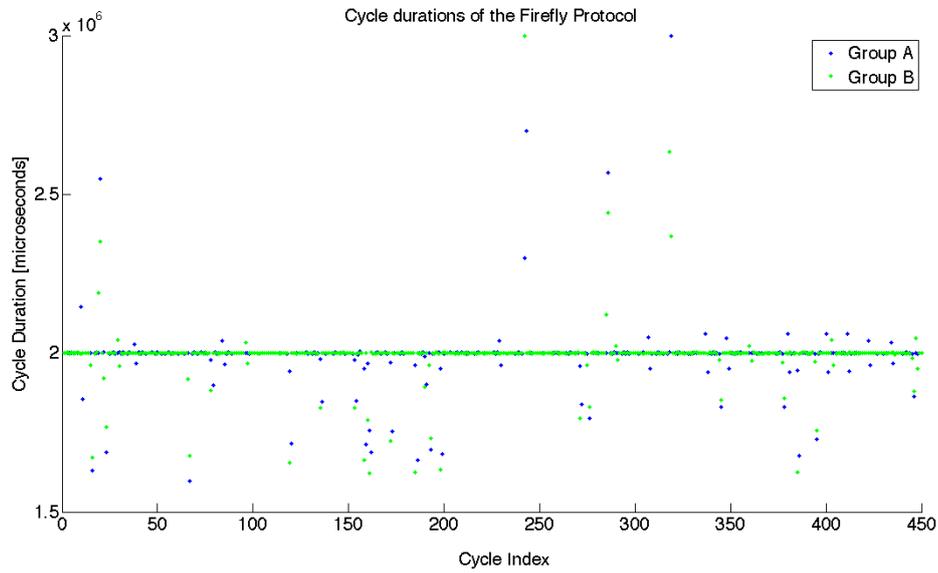

(a) Cycle durations including all outliers. Errors usually occur in the form of shorter cycles.

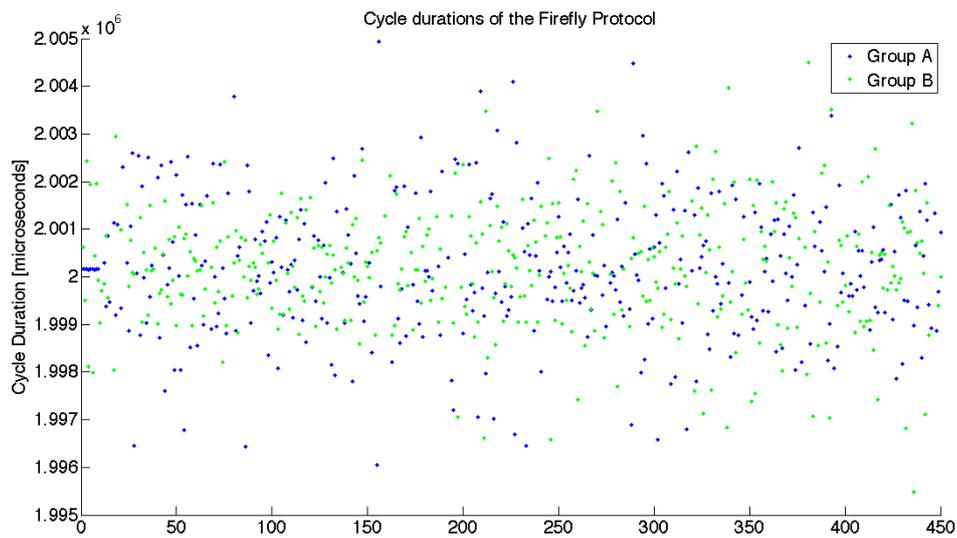

(b) Cycle durations inspected around the mean. The short blue string for the first couple of cycles exists, because the 28 nodes were manually switched on over the course of about 10 cycles. Therefore, for the first 10 cycles not all 28 nodes were part of the system and cycle durations were more precise.

Figure 6.11: Cycle durations in the 28 node Firefly system

# Chapter 7

# Summary and Outlook

In this thesis, the biologically inspired Firefly Protocol (as suggested by Tyrrell et al. (19)) is implemented on MICAz network nodes in TinyOS 2.0. The Firefly Protocol by Tyrrell et al., as well as the "Firefly 2.0" Protocol implemented in this thesis, leave the traditional hierarchical approach in synchronization behind and stride to enable a synchronized network of equal nodes. All synchronization protocols that have been implemented so far have used notions of hierarchy (either static or elected) to clarify the source of reference time in a network (11; 18; 12). The Firefly approach drops this necessity by allowing nodes to reach a consensus as pulse-coupled oscillators. After the system had been implemented, it has been tested on networks consisting of 1, 2, 4, 5 and 28 nodes.

Several adjustments were necessary to apply the theoretical model in practice: The assumption of a constant and predictable transmit time in network nodes does not hold due to internal non-deterministic delays. In addition, the wireless channel requires an access scheme which was not considered in the model. Both of these delays sources can only be compensated a posteriori by the receiver of a synchronization message through knowledge of timestamps, which are created at message creation, transmission and reception. To reach microsecond precision it is necessary to methodically a) compensate for all non-deterministic and b) predict all deterministic delay sources in the system. In addition, this thesis proposes a solution to the dead lock situation which can occur in a sparse Firefly network.

Nodes were found to make mistakes creating large offsets up to 400 ms, and thus, as an effect of the consensus system, irritating the network as a whole. Therefore, the precision of the entire network is of the order of 300 $\mu s$ for five nodes and 2 ms for 28 nodes. However, in a single hop four node network individual nodes can reach up to 1.5 $\mu s$ average offset. Therefore, in matters of precision, this protocol is in the range of the hierarchical approaches. Yet, it remains to be tested over multiple hops. In the current setting, the protocol is not consistent enough, to allow for longer cycle durations which are necessary for better energy efficiency. In regard to the other synchronization protocols which were introduced in the literature survey, this approach is a P2P internal slot synchronization protocol, which requires limited MAC layer access and is realizable on commodity hardware nodes. It is inherently robust, since outliers and dropped packets do not affect the system in the long run.

| Network Size | 4 nodes | 5 nodes | 28 nodes |
|---|---|---|---|
| Average Offset for all nodes | 3 $\mu s$ | 327 $\mu s$ | 1994 $\mu s$ |

Table 7.1: Summary of obtained precisions

The seemingly random large offsets, which are too far from the reference to be simple clock errors, could at this point not be explained. The Firefly system contains dynamical behavior as the belt-like offsets in the five node system. Finding the reason behind this phenomenon should greatly increase timing reliability.

Future work in this field will continue eliminating error sources. To counteract the effects of clock drift, internal clock correction can be added to enable nodes to recognize their own drift from the network reference. This will eliminate the constant disagreement of a couple of microseconds between the nodes. It can be combined with an internal sensation of the node, regarding its own "correctness". It is a strength of the system, that the cycle durations are predefined and fixed. Using this information, a node which has been running with a certain consistence may judge an incoming synchronization message using past experience. Only if it is judged as relevant or is received repeatedly, it will be used. The combination of these methods is expected to combat the large outliers that occur in the present system.

To relieve the system of the slot synchronization limitation, it may possible to enable a posteriori global time perception. Combined with the reliability estimation scheme, a node can count the number of reliable cycles which have passed. This results in a time frame which can be attached to measurement results. The data collector can then infer backwards when a measurement was taken, thus creating the notion of global time knowledge.

Regarding energy efficiency, the network scheme must include sleep modes. Optimally, the cycle length of the Firefly Protocol could be increased to the range of hours, thus allowing extensive sleeping periods to maximize battery lifetimes.

# Appendix A

# Appendix

## A.1  Tools

At many points during this work, it was necessary to check, evaluate or measure components of a node or the entire system. For this purpose, various test applications were created which are shortly mentioned here and can be found on the accompanying CD.

### A.1.1  BaseStation

The BaseStation component was the data collector for the Firefly network. The standard BaseStation which is provided with TinyOS was modified to have an increased message buffer. All incoming radio messages are timestamped and passed to the serial port.

### A.1.2  BSJitterSerial

The original base station jitter measurement component was used to measure radio sending hardware delays and medium access delays. It was extended over the course of the project to perform outgoing and incoming message timestamps, AM_ID-dependent timestamps and shut-off of CCA. Later it was the basis for the round trip measurement system as described in Section 4.2.2.

### A.1.3  LabSync

LabSync was the most basic approach to a firefly system and was designed as a problem submission for the wireless sensor laboratory. Operating in the millisecond range, LabSync allows to blink LEDs synchronously.

### A.1.4 ReadTime

The RadioTimeStamping interface was explored in the ReadTime component. As described in Section 4.2.2, the 8-bit counter was found to be too narrow.

### A.1.5 TimerCheck

The TimerCheck component was used to compare the different timers that are available on MICAz. By starting, stopping and restarting various timers repeatedly, hardware delays and timer reliability were explored.

### A.1.6 TrafficSource

A mote which runs TrafficSource will continually send back-to-back messages. Each successful transmission of a dummy packet triggers the transmission of the next message. It was used to push the Contention Backoff of the CSMA scheme.

## A.2 TinyOS Modules and Components Used

The most straightforward way to describe the components of the Firefly Protocol is to print the contents of the configuration file. All components and wirings used are listed below:

```
// Main component
components FireflyC as App;

components MainC;
App.Boot -> MainC;

// LED control
components LedsC;
App.Leds -> LedsC;

// Random number generator (used in the dead lock circumvention)
components RandomC;
App.Random -> RandomC;

// Timers
components CounterMicro32C as Counter32;
App.Counter -> Counter32;
```

```
components new AlarmMicro32C() as Alarm32;
App.Alarm -> Alarm32;

// ActiveMessageRadio
components ActiveMessageC as AM;
App.RadioControl -> AM;
App.RadioPacket -> AM;
App.SyncSend -> AM.AMSend[AM_FIREFLY_SYNC_MSG];
App.SyncReceive -> AM.Receive[AM_FIREFLY_SYNC_MSG];

// Radio chip control interface - used for timestamping
components CC2420TransmitC;
App.RadioTimeStamping -> CC2420TransmitC;
App.CC2420Transmit -> CC2420TransmitC;

// Optional interface for backoff modification and CCA control
components CC2420CsmaC;
App.RadioBackoff -> CC2420CsmaC.RadioBackoff[AM_FIREFLY_SYNC_MSG];
```

## A.3   Errors

**Node Lock-up**

There exists an error where the network node software freezes. The frozen node will remain with one LED switched on and not be susceptible to incoming messages. It is assumed that this problem is caused by the message receive interrupt. While one synchronization message is still being processed and states are about to change, the next message interrupt is triggered, thus stopping the state change and locking the node. This error has not been investigated further as it can be easily recognized in the laboratory and fixed through a reboot. It becomes more probable with growing network size and has been not encountered with less than 5 nodes. The probability of this error is roughly estimated at 1 lock-up per 5,000 synchronization messages (e.g. 10 nodes going through 500 cycles will probably cause one lock-up).

# List of Figures

# List of Tables